\documentclass[a4paper,10pt,aps,pre,longbibliography,twocolumn,amsmath,amssymb,nofootinbib,superscriptaddress]{revtex4-2}
\usepackage{graphicx}

\usepackage[colorlinks=true,linkcolor=blue]{hyperref}
\usepackage{amsmath, bm}

\DeclareMathAlphabet{\mathitbf}{OML}{cmm}{b}{it}

\newcommand{\rv}{\mathitbf r}
\newcommand{\xv}{\mathitbf x}

\newcommand{\Gv}{\mathitbf G}

\newcommand{\mathBold}[1]{\mbox{\boldmath$#1$}}
\newcommand\sbullet[1][.5]{\mathbin{\vcenter{\hbox{\scalebox{#1}{$\bullet$}}}}}

\begin{document}

\title{Predicting plasticity in disordered solids from structural indicators}

\author{D.~Richard}
\affiliation{Institute for Theoretical Physics, University of Amsterdam, Science Park 904, Amsterdam, the Netherlands}
\affiliation{Department of Physics, Syracuse University, Syracuse, New York 13244}
\author{M.~Ozawa}
\affiliation{Laboratoire de Physique de l’Ecole Normale Supérieure, ENS, Université PSL, CNRS, Sorbonne Université, Université de Paris, F-75005 Paris}
\affiliation{Laboratoire Charles Coulomb, UMR 5221 CNRS-Université de Montpellier, Montpellier, France}
\author{S.~Patinet}
\affiliation{PMMH, CNRS UMR 7636, ESPCI Paris, PSL University, Sorbonne  Université, Université  de Paris, F-75005 Paris, France}
\author{E.~Stanifer}
\affiliation{Department of Physics, Syracuse University, Syracuse, NY 13244}
\author{B.~Shang}
\affiliation{Beijing Computational Science Research Center, Beijing 100193, China}
\affiliation{Univ. Grenoble Alpes, CNRS, LIPhy, 38000 Grenoble, France}
\author{S.A.~Ridout}
\affiliation{Department of Physics and Astronomy, University of Pennsylvania, Philadelphia, Pensylvania 19104, USA}
\author{B.~Xu}
\affiliation{Beijing Computational Science Research Center, Beijing 100193, China}
\affiliation{Mechanical Engineering, Johns Hopkins University, Baltimore, Maryland 21218, USA}
\author{G.~Zhang}
\affiliation{Department of Physics and Astronomy, University of Pennsylvania, Philadelphia, PA 19104}
\author{P.K.~Morse}
\affiliation{Department of Chemistry, Duke University, Durham, North Carolina 27708, USA}
\author{J.-L.~Barrat}
\affiliation{Univ. Grenoble Alpes, CNRS, LIPhy, 38000 Grenoble, France}
\author{L.~Berthier}
\affiliation{Laboratoire Charles Coulomb, UMR 5221 CNRS-Université de Montpellier, Montpellier,  France}
\affiliation{Department of Chemistry, University of Cambridge, Lensfield Road, Cambridge CB 2 1EW, United Kingdom
}
\author{M.L.~Falk}
\affiliation{Materials Science and Engineering, Johns Hopkins University, Baltimore, Maryland 21218, USA}
\affiliation{Mechanical Engineering, Johns Hopkins University, Baltimore, Maryland 21218, USA}
\affiliation{Physics and Astronomy, Johns Hopkins University, Baltimore, Maryland 21218, USA}
\affiliation{Hopkins Extreme Materials Institute, Johns Hopkins University, Baltimore, Maryland 21218, USA}
\author{P.~Guan}
\affiliation{Beijing Computational Science Research Center, Beijing 100193, China}
\author{A.J.~Liu}
\affiliation{Department of Physics and Astronomy, University of Pennsylvania, Philadelphia, PA 19104}
\author{K.~Martens}
\affiliation{Univ. Grenoble Alpes, CNRS, LIPhy, 38000 Grenoble, France}
\author{S.~Sastry}
\affiliation{Jawaharlal Nehru Center for Advanced Scientific Research, Jakkur Campus, Bengaluru 560064, India}
\author{D.~Vandembroucq}
\affiliation{PMMH, CNRS UMR 7636, ESPCI Paris, PSL University, Sorbonne  Université, Université  de Paris, F-75005 Paris, France}
\author{E.~Lerner}
\affiliation{Institute for Theoretical Physics, University of Amsterdam, Science Park 904, Amsterdam, Netherlands}
\author{M.L.~Manning}
\affiliation{Department of Physics, Syracuse University, Syracuse, NY 13244}

\begin{abstract}

Amorphous solids lack long-range order. Therefore identifying structural defects --- akin to dislocations in crystalline solids --- that carry plastic flow in these systems remains a daunting challenge. By comparing many different structural indicators in computational models of glasses, under a variety of conditions we carefully assess which of these indicators are able to robustly identify the structural defects responsible for plastic flow in amorphous solids. We further demonstrate that the density of defects changes as a function of material preparation and strain in a manner that is highly correlated with the macroscopic material response. Our work represents an important step towards predicting how and when an amorphous solid will fail from its microscopic structure.

\end{abstract}

\maketitle

\section{Introduction}
How can we predict when and where a material will fail? For disordered solids, including many food and cosmetic products, screens and cases for smartphones, and even mud and gravel perched on a hillside, this fundamental question remains a challenge. Under small deformations or forces, such amorphous materials respond as an elastic solid, but beyond a critical threshold the materials yield and exhibit extensive irreversible plastic deformation.

At the moment, we cannot easily predict from first-principles whether a given material will fail abruptly and catastrophically, termed brittle failure, or flow slowly and steadily, known as ductile flow. Moreover, we can not predict when or where it will fail, and we lack global design principles for how we might change the microscopic structure of these materials in order to control failure mechanisms.

One reason disordered solids are so challenging to understand is that, unlike crystals, their microscopic structure lacks long-range order. In crystals it is easy to identify a defect where the crystalline order is broken, and unsurprisingly plasiticity is initiated at certain types of these defects. Over the past 50 years, analogous structural defects have been proposed in amorphous solids~\cite{spaepen1977microscopic}, but it has proven more difficult to identify these and connect them to deformation and failure.

For this reason, theoretical work has largely remained disconnected from simulations. Several fairly successful theories have posited that there is no correlation of particle rearrangements with the inherent microstructure~\cite{ritort2003glassy}, while others such as Shear Transformation Zone (STZ)~\cite{argon1979plastic, falk1998dynamics}, Soft Glassy Rheology (SGR)~\cite{sollich1997rheology} and elastoplastic models~\cite{nicolas2018deformation} assume the existence of structural defects, without specifying the precise definition of these defects. Perhaps more problematically, all these theories are phenomenological in the sense that they contain fitting parameters that we do not know how to extract from first principles, i.e. from the microstructure and interactions between the constituent particles of the material. In fact, even in the case of crystalline solids, predicting the collective dynamics of many dislocations~\cite{miguel2001intermittent,salmenjoki2020plastic} or dislocation nucleation in defect free single crystals remain open problems \cite{van2003quantifying,miller2008nonlocal,garg2016mechanical,zhang2017taming,luo2019plasticity,reddy2020nucleation}. Hence, tools developed to characterize amorphous solids can serve as well to understand plasticity in crystalline solids \cite{rottler2014predicting,ovaska2017excitation,sharp2018machine}.

Fairly recently, large-scale computer simulations of glass-formers have reinvigorated the search for structural defects. Examples of promising approaches include identifying energetically favored structures and community inference~\cite{coslovich2007understanding,fang2010atomistic,keys2011characterizing,malins2013identification,tong2018revealing,shirai2019microscopic,wei2019assessing,paret2020assessing}, mapping the local shear modulus~\cite{tsamados2009local}, highlighting regions that are excited by linear and nonlinear vibrational modes~\cite{widmer2008irreversible,tanguy2010vibrational, manning2011vibrational, gartner2016nonlinear,zylberg2017local,tong2014order,schwartzman2019anisotropic,xu2019atomic}, or quantifying the structure in more complex ways e.g.~using machine learning with either supervised~\cite{cubuk2015identifying,schoenholz2016structural,bapst2020unveiling} or unsupervised methods~\cite{ronhovde2011detecting,boattini2019unsupervised,boattini2020autonomously}. A particularly literal and successful approach applies strain to small regions within a simulation to determine which are closest to yielding~\cite{patinet2016connecting,patinet2019origin}. 

Until now there have been three main drawbacks to this line of inquiry.  First, there has never been a consistent methodology for evaluating whether a given indicator that identifies structural defects works well for predicting deformation and failure, although efforts towards this goal have been made~\cite{patinet2016connecting}.  Second, it has not been clear whether a given method works best only on a particular model system or the interaction potential for which it was designed, or whether some methods work well universally across different disordered solids. Finally, computer-generated amorphous solids have historically been vastly more ductile than those in real experiments, and so it was difficult to simulate bulk brittle materials that exhibit catastrophic failure. Recent methodological developments~\cite{berthier2016equilibrium,ninarello2017models} now allow very deep supercooling of model polydisperse liquids, to temperatures comparable to and even lower than what can be achieved with laboratory liquids. These liquids, when quenched to zero temperature, form ultra-stable glasses that exhibit brittle failure~\cite{ozawa2018random}. In this paper we use the term brittle to characterize a discontinuous yielding characterizing the mechanical instability associated with the formation of a shear band. Although this phenomenon is not accompanied by the formation of free surfaces, as seen in the fracture of brittle materials, the macroscopic avalanche taking place at the discontinuous yielding transition would be the precursor of a crack in the absence of periodic boundary conditions employed in simulations.

In this article we employ these computational tools to conduct a comprehensive and quantitative comparative study of how well several recently-proposed structural indicators predict plastic activity in two-dimensional model glasses formed using a wide range of preparation conditions. We develop a standard methodology for comparing these indicators to one another, and to the complex deformation fields that result from the applied deformation.

We find that different classes of structural indicators are not always strongly correlated, suggesting that different paradigms for identifying structural defects are sensitive to distinct structural information. With a few exceptions, the indicators we investigated are excellent at predicting the loci of plastic instabilities in ductile and brittle materials over short strain scales (0.1\% in the system sizes studied), and several remain correlated beyond 10\% strain, highlighting that structure does indeed govern plastic deformation in these zero-temperature materials under simple shear. The quality of a given indicator does not vary much between the two interaction potentials we studied. In contrast, their predictive capabilities do change with glass stability: indicators are generally less accurate in ductile glasses, where many regions are soft even at zero strain. An exception to this rule occurs in the ultra-stable glasses accessible via swap Monte Carlo -- in these materials many of the structural indicator fields change so much before the first plastic rearrangement that the indicator field at zero strain is not highly predictive. Importantly, we demonstrate that free volume (or any measure of a local density), historically used to predict plasticity in metallic glasses, performs much worse than other indicators. Finally, we are able to follow the complete strain history of a given brittle sample and we find that our indicators are able to capture an anisotropic spatial distribution of soft regions before shear banding.

\begin{figure*}[t!]
  \includegraphics[scale=1.1]{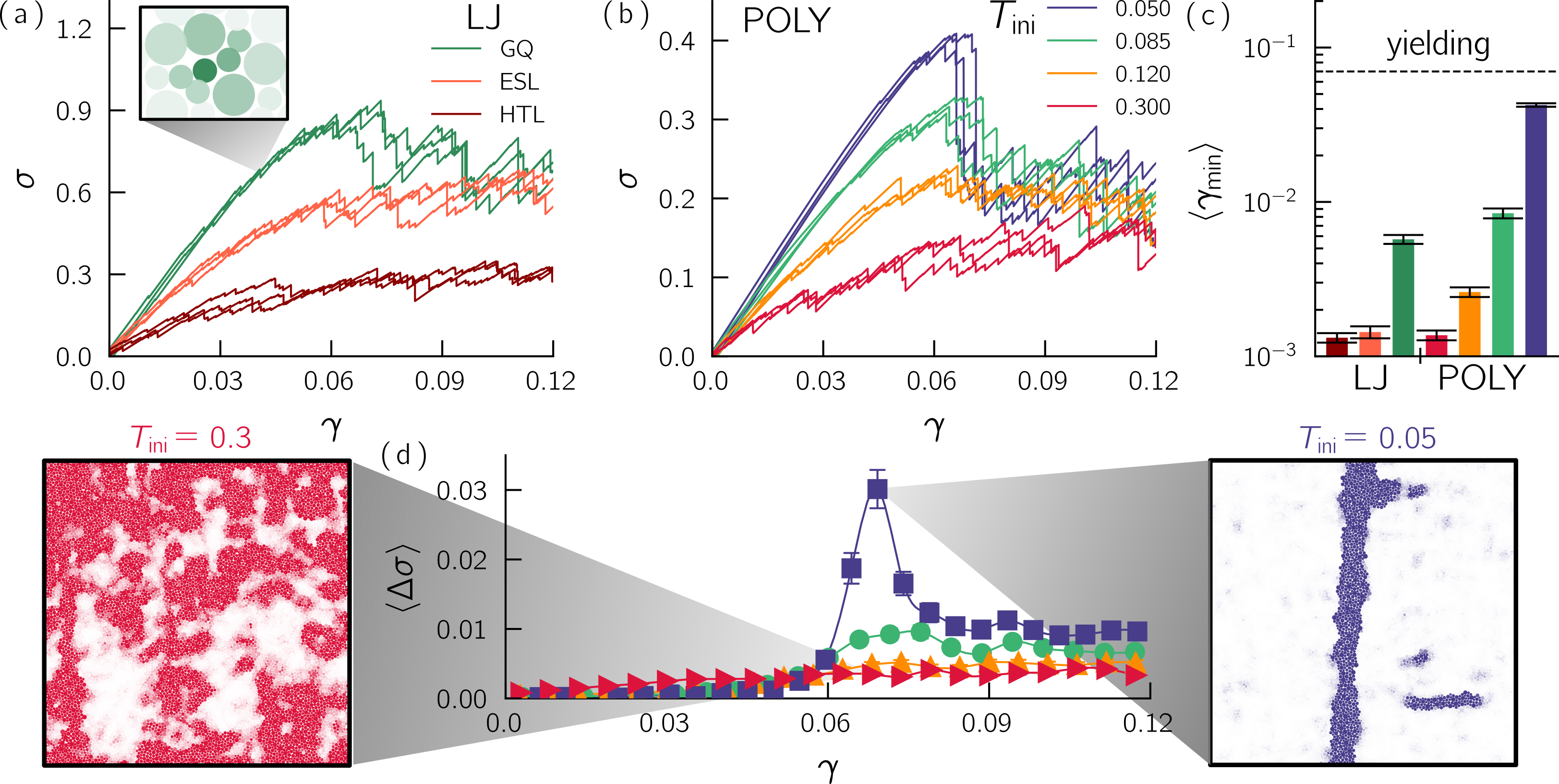}
  \caption{\textbf{Probing plasticity from ductile to brittle glasses.} (a) Typical stress-strain curves for the LJ system prepared via three different quench protocols. Inset shows a typical plastic event where dark particles indicate nonaffine rearrangements. (b) Typical stress-strain curves for the POLY system quenched from four initial temperatures $T_{\text{ini}}$. (c) Average strain at the first plastic instability $\langle\gamma_{\text{min}}\rangle$ for the different systems and protocols (with $N=10^4$ particles). The dashed line indicates the yield strain $\gamma_{\text{yielding}}\simeq7\%$. (d) Average stress drop $\langle\Delta\sigma\rangle$ as a function of the strain $\gamma$ for the POLY system. Snapshots show the cumulative nonaffine displacements observed after yielding in a very stable glass (bottom right) and a typical ductile glass quenched instantaneously from a high temperature liquid (bottom left). Errorbars are sample-to-sample fluctuations.}
  \label{fig:1}
\end{figure*}

\section{Method}
\subsection{From ductile to brittle materials}
We shear systems of 2D soft disks using a standard athermal quasistatic (AQS) protocol, where the only control parameters are the imposed strain $\gamma$ and the strain increment $\delta\gamma$ \cite{maloney2006amorphous}; see Appendix~\ref{ap:protocol} for details. Fig.~\ref{fig:1}(a-b) shows that the mechanical response quantified by plots of the shear stress $\sigma$ versus $\gamma$, where different curves correspond to different material preparation protocols. Fig.~\ref{fig:1}(a) corresponds to a standard glass former, a bidisperse packing of particles with a Lennard-Jones (LJ) interaction potential; glasses were formed by cooling equilibrium states by means of conventional molecular dynamics methods~\cite{barbot2018local}. Fig.~\ref{fig:1}(b) corresponds to polydisperse (POLY) disks~\cite{berthier2016equilibrium} interacting via purely repulsive interactions; liquid states spanning a wide range of supercooling temperatures were equilibrated using the swap Monte Carlo method, and quenched to the glassy phase with a minimization algorithm (details about models and protocols are provided in Appendix~\ref{ap:protocol}). Plastic instabilities associated with deformation and particle rearrangements correspond to instantaneous drops in the stress. Qualitatively, we see that some material preparation protocols generate less-stable, ductile solids where small plastic instabilities occur frequently until the system approaches a steady state. Other preparation protocols generate ultra-stable brittle solids which behave nearly elastically until they fail catastrophic around 6 or 7 \% strain ($\gamma_{\text{yielding}}\!\simeq\!0.07$ ). In order to compare these two very different material systems, we study the average strain at the first instability, $\langle\gamma_{\text{min}}\rangle$ (Fig.~\ref{fig:1}(c)), which is a good indicator for material stability/ductility for a given finite system size $N$, and correlates strongly with other previously developed measures for stability (Supplementary Fig. S1). Another indicator, the magnitude of the average stress drop $\langle\Delta \sigma\rangle$, is an order parameter for the macroscopic yielding transition~\cite{ozawa2018random}, exhibiting a sharp peak at the macroscopic yielding transition in brittle systems (Fig.~\ref{fig:1}(d)) associated with the formation of a system-spanning shear band (right inset).

\subsection{Structural indicators for plastic defects}

In this paper, we have considered $18$ different indicators sorted into five different families:
\begin{itemize}
\item (i) Eight purely structural ones that require only the position of particles and not their interaction potential (red)
\item (ii) One machine learning-based method that is also structural but require a training on a subset of shear deformations (purple)
\item (iii) Three based on the linear response/harmonic vibrational modes with no information about the applied strain (green)
\item (iv) Three that quantify the linear response to a specialized applied strain, which here is simple shear (blue)
\item (v) Three going beyond the linear response (orange).
\end{itemize}
We also have considered two additional indicators (colored in black) that are not \textit{a priori} related to plasticity (the local potential energy $\varphi$ and local thermal expansion $\alpha$). Detailed descriptions of all employed structural indicators in this work are provided in Appendix~\ref{ap:indicators}. In Fig.~\ref{fig:1b}, we provide schematics explaining the key physical ingredients behind the calculation of the different classes defined above.

\begin{figure*}[t!]
  \includegraphics[scale=0.95]{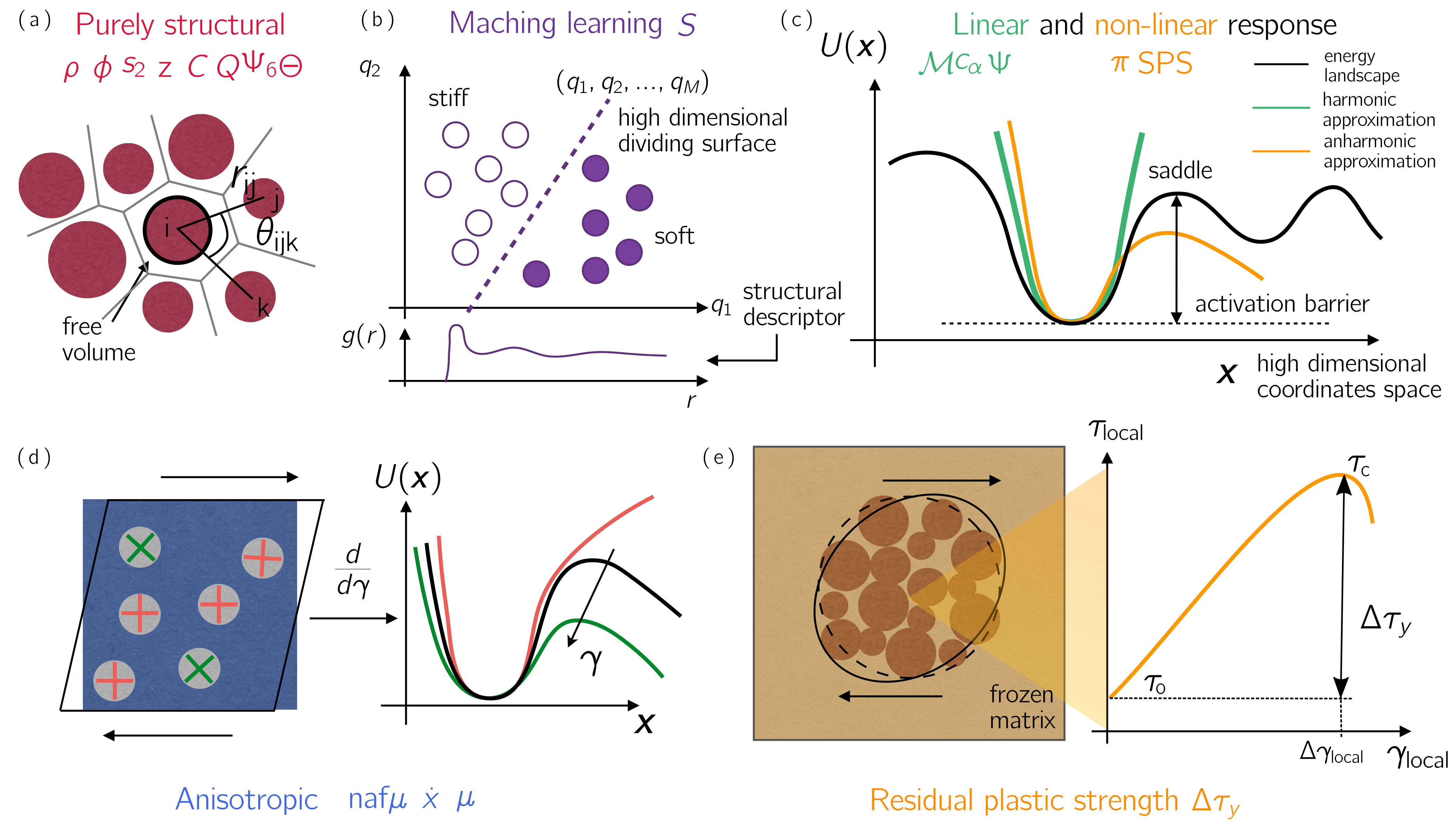} 
  \caption{\textbf{Different classes of structural indicators.}~(a) Purely structural indicators~\cite{kawasaki2007correlation,rieser2016divergence,piaggi2017entropy,tong2018revealing} that require only information about particle positions or geometry such as the distance between a particle pair $r_{\rm ij}$ or angle between a triplet $\theta_{\rm ijk}$. These methods enable the detection of locally unstable structures responsible for soft spots. (b) Machine learning-based methods \cite{cubuk2015identifying,ronhovde2011detecting,boattini2019unsupervised,bapst2020unveiling} correlate a high-dimensional description of a local particle environment (constructed from purely structural information) with the likelihood of undergoing a local shear transformation. The cartoon explains the Softness method~\cite{cubuk2015identifying,schoenholz2016structural} that utilizes Support Vector Machine to distinguish "soft" and "stiff" particles based on their local radial distribution function $g(r)$ (counting the number of neighbors at a distance $r$). (c) Sketch of a rough energy landscape around an inherent state of the glass. The green and orange lines are a harmonic and anharmonic approximation of the landscape, respectively. Linear response methods \cite{widmer2008irreversible,tanguy2010vibrational,manning2011vibrational,zylberg2017local,tong2014order} associate soft regions to low curvatures of the energy landscape. Anharmonic methods can provide estimates of activation barriers \cite{lerner2016micromechanics,xu2017strain,xu2018predicting}. (d) Anisotropic indicators are also based on the linear response of the system but include the geometry of the imposed deformation \cite{maloney2006amorphous,tsamados2009local,schwartzman2019anisotropic,xu2019atomic}. An explicit derivative of quantities with respect to the imposed strain $\gamma$ allows filter out soft spots that couple weakly (red crosses) with the shear geometry. Soft regions with the correct shear orientation (green crosses) will correspond to activation barriers that decrease upon incrementation of the strain (softening shown in green). (e) Schematic of the residual plastic strength $\Delta \tau_y$ from the frozen matrix method \cite{patinet2016connecting,patinet2019origin}. A local patch of the material (with prestress $\tau_0$) is sheared until it yields (at a local plastic strength $\tau_c$), which allows the extraction of an estimate for the local strain distance to threshold $\Delta\gamma_{\rm local}=\Delta\tau_y/\tilde{\mu}$, where $\Delta\tau_y=\tau_c-\tau_0$ and $\tilde{\mu}$ denote the minimal residual plastic strength and bulk shear modulus, respectively.}
  \label{fig:1b}
\end{figure*}

\subsection{Rank correlation between structure and plastic events}
Having characterized the macroscopic material properties of these systems, we next study how plastic deformation correlates with proposed indicators for structural defects. Although many different correlation functions have been proposed in previous work, here we adopt a very simple cumulative rank correlation originally put forward in Ref.~\cite{patinet2016connecting}, in order to fairly compare structural indicators with vastly different magnitudes and distributions, and to avoid setting an arbitrary threshold.

\begin{figure}[t!]
  \includegraphics[scale=1.]{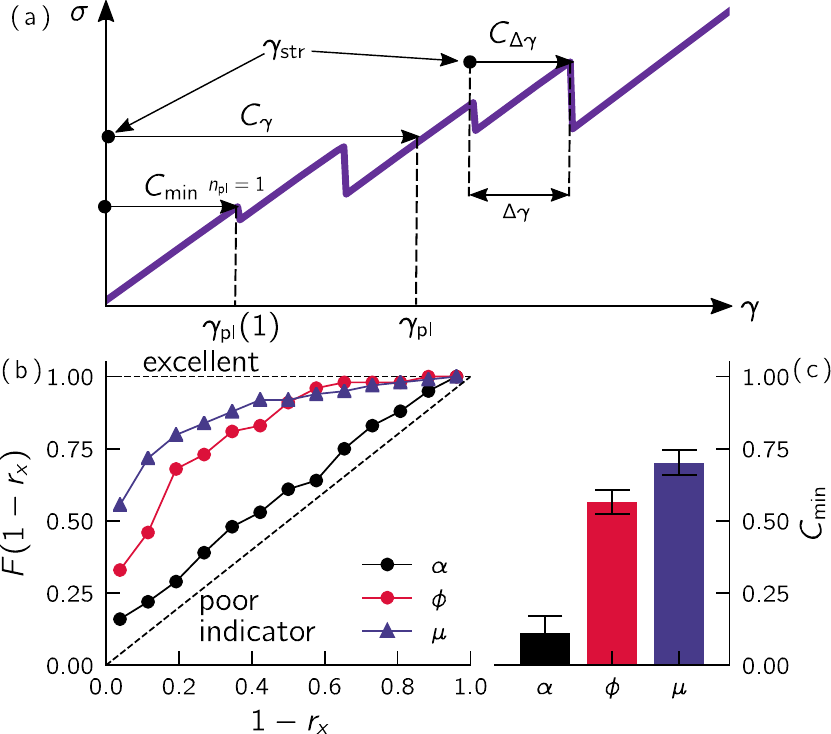}
  \caption{\textbf{Rank correlation.} (a) Sketch of the different correlations $C_{\rm min}$, $C_\gamma$, and $C_{\Delta \gamma}$ considered in this work. The bullet in each arrow marks the strain $\gamma_{\rm str}$ at which the structural field is computed and the head's arrow points to the event at which it is compared. (b) Cumulative distribution $F(1-r_x)$ of the rank $r_x$ at the first plastic instability for the thermal expansion $\alpha$, free volume $\phi$, and local shear modulus $\mu$. (c) Correlation score $C_{\text{min}}$ at the first plastic instability.}
  \label{fig:3}
\end{figure}

In particular, we ask how well a structural field is able to capture the triggering event associated with a plastic instability. The latter is detected by computing the destabilizing critical mode $\boldsymbol{\Psi_c}$ at the onset of each event. From this methodology, we define $C=C(\gamma_{\rm str},\gamma_{\rm pl})$ as the correlation between a snapshot of the structure at strain $\gamma_{\rm str}$, and the plastic deformation field at a strain $\gamma_{\rm pl}$ (measured by $\boldsymbol{\Psi_c}$). We consider three different prediction scores, sketched in Fig.~\ref{fig:3}(a): (i) $C_{\text{min}}$ correlating a structural field computed at zero strain $\gamma_{\rm str}=0$ with the first plastic instability occurring at $\gamma_{\rm pl}(1)$, (ii) $C_\gamma$ scoring how well a structural field computed at $\gamma_{\rm str}=0$ is able to predict events located around $\gamma_{\rm pl}$ (or alternatively the $n$'th plastic event), and finally (iii) $C_{\Delta\gamma}$ is defined as the correlation between a structural field at a strain $\Delta\gamma>0$ prior to a plastic event, and the subsequent plastic event.

In practice, we rank-order the value of each indicator on each particle between $1$ and $N$. We choose a larger rank to be associated with a higher probability of rearranging, e.g small shear moduli or large free volumes. We define $r_x (\gamma_{\rm str}, \gamma_{\rm pl})$ as the rank of the structural indicator at a strain $\gamma_{\rm str}$ on the particle associated with the largest $|\boldsymbol{\Psi_c}|$ evaluated at a strain $\gamma_{\rm pl}$, normalized by the number of particles. For a good structural indicator, we expect $r_x$ to be nearly unity. Sampling $r_x$ over many realizations (here $100$ independent samples), we expect the cumulative distribution function $F(1-r_x)$ to approach a Heaviside step function or a linear behavior $F(1-r_x)\sim 1-r_x$ for an excellent and poor indicator, respectively. Figure~\ref{fig:3}(b) illustrates this correlation for selected structural indicators: the local shear modulus $\mu$, the free volume $\phi$, and the thermal expansion $\alpha$. As one would expect, we find no sign of correlation between thermal expansion and plasticity, but we do observe that $\mu$ and $\phi$ correlate with plastic rearrangements. The cumulative rank correlation between a snapshot of the structure at strain $\gamma_{\rm str}$ and the plastic deformation field at a strain $\gamma_{\rm pl}$ is then defined as $C=2\langle{r_x}\rangle-1$~\cite{patinet2016connecting}, which ranges from $0$ (poor indicator) to $1$ (excellent indicator) (see example in Fig ~\ref{fig:3}(c)). Note that $\langle r_x\rangle$ simply corresponds to the integral of $F(1-r_x)$ between $0$ and $1$.

In addition, we compute the degree of similarity between two structural indicators $A$ and $B$ as a cross rank Spearman correlation
\begin{equation}
\label{eq:spearman}
C_s=1-6\sum_i^N (a_i-b_i)^2/(N(N^2-1)),
\end{equation}
where $a_i$ and $b_i$ are the rank of particle $i$ for the metric $A$ and $B$, respectively.

\begin{figure*}[t!]
  \includegraphics[width = \textwidth]{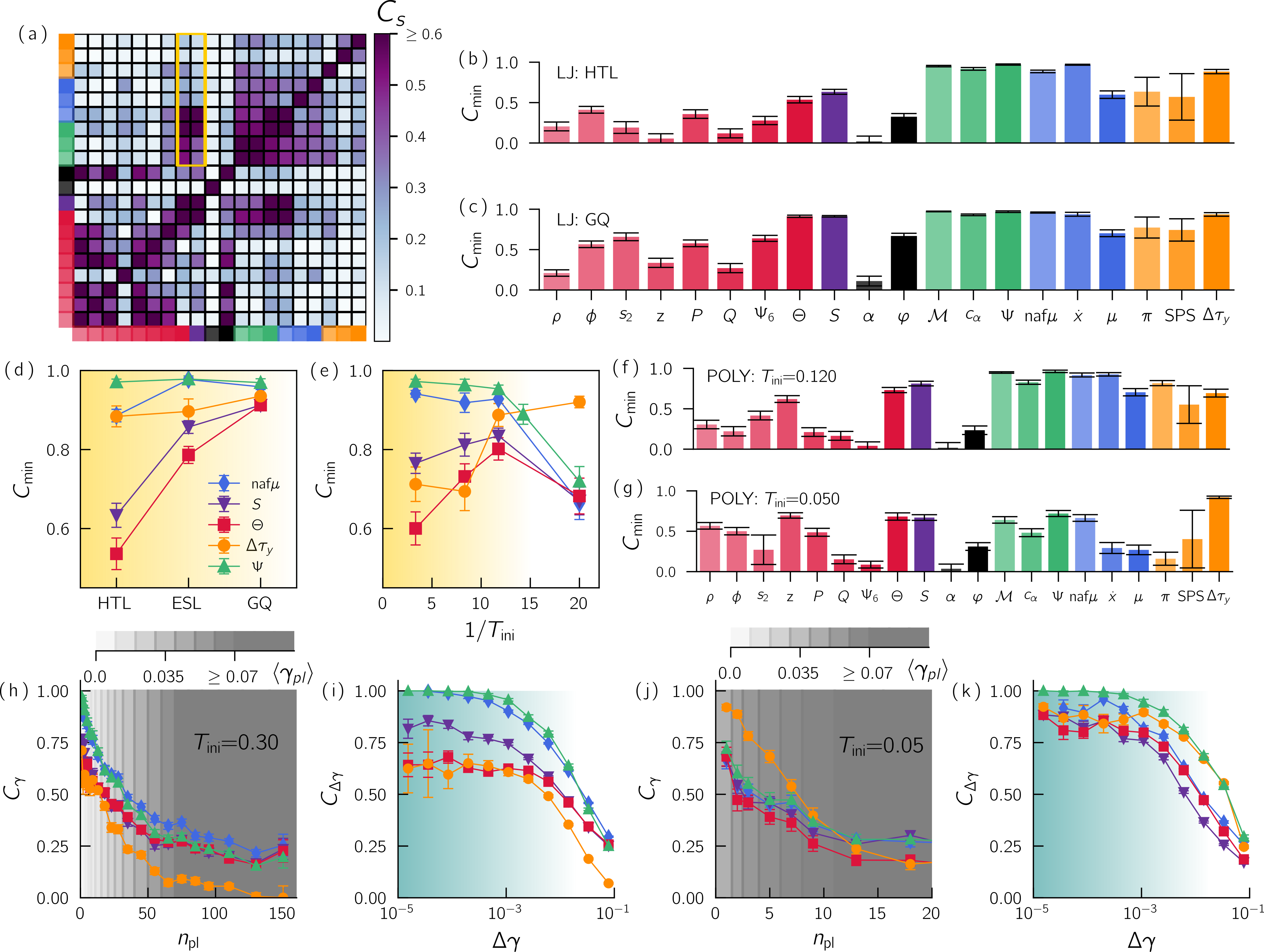}
  \caption{\textbf{Comparison of structural indicators.}~ Local density $\rho$, free volume $\phi$, excess entropy $s_2$, contact force number $z$, voronoi anisotropy $P$, divergence of the voronoi anisotropy $Q$, hexatic bond orientational order $\Psi_6$, steric bond order $\Theta$, softness field $S$, local thermal expansion $\alpha$, potential energy $\varphi$, low-frequency vibrational modes $\mathcal{M}$, local heat capacity $c_\alpha$, vibrality $\psi$, atomic nonaffine shear modulus $\text{naf}\mu$, nonaffine velocity $\dot{x}$, local shear modulus $\mu$, nonlinear vibrational modes $\pi$, Saddle Point Sampling $\text{SPS}$, residual plastic strength $\Delta\tau_y$. (a) Indicator to indicator Spearman cross correlation $C_s$ for the less ductile (GQ) LJ glasses. (b-c) and (f-g) Correlation $C_{\text{min}}=C(0, \gamma_{\rm pl}(1))$ between a structural field computed at $\gamma=0$ and the first plastic event. Binary LJ data are shown for the most (HTL) to the least (GQ) ductile glasses in (b) and (c), respectively. Panel (d) shows the same information as a function of the glass stability. Polydisperse glasses data prepared by SWAP are shown for mildly ($T_{\text{ini}}=0.12$) and very stable glasses ($T_{\text{ini}}=0.05$) in (f) and (g), respectively. Panel (e) shows the same information as in (d) but plotted as a function of the inverse of the parent temperature. (h,j) Correlation decay $C_\gamma=C(0,\gamma_{\rm pl}(n_{\rm pl}))$ between a structural field computed at $\gamma=0$ and the \textit{nth} plastic event in a system composed of $10^4$ particles. The underlying gray gradient indicates the corresponding average plastic strain $\langle \gamma_{\rm pl} \rangle$ ranging linearly from $\gamma=0$ (transparent) to $\gamma\ge\gamma_{\text{yielding}}$ (opaque). (i,k) Correlation growth $C_{\Delta \gamma}=C(\gamma_{\rm pl}-\Delta\gamma,\gamma_{\rm pl})$ between a structural field $\Delta\gamma$ away from a plastic event located at $\gamma_{\rm pl}$. Results are for ductile glasses (h-i) and brittle glasses (j-k) prepared at $T_{\text{min}}=0.30$ and $T_{\text{min}}=0.05$, respectively. The different colors and symbols correspond to the same structural indicators as shown in (d) and (e).}
  \label{fig:4}
\end{figure*}

\section{Correlation of structural defects with plasticity}

We start by comparing the spatial similarities between the whole set of structural indicators considered. The full Spearman indicator to indicator correlation $C_s$ is shown in Fig.~\ref{fig:4}(a) for the gradually quenched binary Lennard-Jones system. The majority of purely structural indicators correlate strongly with each other but are not strongly correlated with indicators based on linear response. This suggests that the two classes of indicators capture different structural features, likely because the latter has explicit access to the forces across bonds, which encode longer-range elastic interactions across the contact network.

\subsection{Prediction of the first plastic event}
Previous work has demonstrated that a glass' formation protocol has a substantial impact on where and how it fails under shear. Therefore, we first study how the structural indicators evaluated at zero strain, $\gamma=0$, correlate with the plastic deformation that occurs at the first plastic event, at a strain $\gamma_{\rm pl}(1)$. Figure~\ref{fig:4}(b,c) show the cumulative rank correlation $C_{\text{min}}=C(0, \gamma_{\rm pl}(1))$ for all the indicators studied in the most ductile (HTL) and least ductile (GQ) LJ glasses. Note that $C_{\text{min}}$ is evaluated at vastly different strains in different systems, at approximately $\gamma \sim 0.1$\% in ductile glasses, and approximately $\gamma \sim 5$\% in the most brittle glasses. We observe a qualitative agreement between the two different protocols with some quantitative change in the predictiveness as the sample becomes more stable. Indicators constructed from vibrational modes (from $\mathcal{M}$ to $\Psi$) perform extremely well, with a correlation $C_{\text{min}}$ approaching unity. These indicators can even perform as well as the residual plastic strength $\Delta\tau_y$, which is a highly nonlinear method that locally shears a small portion of the material centered around a particle $i$ and measures how much additional stress is needed to induce yielding. This result confirms many observations that loci of plastic instabilities are directly connected to the presence of low-frequency excitations that control the response of a system upon external driving \cite{manning2011vibrational,zylberg2017local}. 

The majority of structure-based indicators perform poorly, with two important exceptions: the steric bond order $\Theta$ and the machine learning-based softness field $S$, which are highly correlated with each other and with soft modes, as shown by $C_s$ in Fig.~\ref{fig:4}(a), highlighted by a yellow box. The indicator $\Theta$ highlights particles that significantly depart from sterically favored configurations with little particle overlap, and thus is a local measure of frustration. The fact that $\Theta$ is so similar to the agnostic machine learning method suggests that the machine learning algorithm has learned to identify such frustration, too. In metallic glasses, plastic rearrangements do correlate strongly with unfavored local structures \cite{ding2014soft}. This connects to recent evidence that internal stresses caused by frustration built-up during the quench process are responsible for quasilocalized excitations \cite{lerner2018frustration} and strongly echoes with experimental work on metallic glasses~\cite{zhang2006making}. The same quasilocalized excitations control the regions with a low residual strength. In particular, recent works~\cite{lerner2016micromechanics,kapteijns2019nonlinear} have demonstrated that their distance to threshold scales with the cube of their frequency, i.e. $\Delta\tau_y\sim \omega^3$ when $\omega\to0$. Together, this establishes a connection between a geometrical signature of frustration (measured by $\Theta$), regions with high internal stresses which create low-frequency modes (measured by $\Psi$), and regions with small distance to threshold (measured by $\Delta\tau_y$).

In Fig.~\ref{fig:4}(d), we summarize our results by plotting $C_{\text{min}}$ for the three different quench protocols shown in Fig.~\ref{fig:1}(a) and ordering them according to the degree of stability of the sample. For visibility, we display only one of the best structural indicators in each family ($\Theta$, $S$, $\mathcal{M}$, naf$\mu$, and $\Delta\tau_y$). The predictive power increases as the system becomes less ductile, consistent with Ref.~\cite{barbot2018local}. In the least ductile LJ systems, structural indicators based on vibrational modes or frustrated geometric configurations work very well to predict the next plastic event. This is especially impressive since less ductile materials must be strained further before triggering the next plastic event.

Figure~\ref{fig:4}(e) shows the correlation $C_{\text{min}}$ for the same structural quantities as in Fig.~\ref{fig:4}(d), now for a polydisperse system prepared using swap Monte Carlo (POLY). Again, the initialization protocols are ranked from most ductile to most brittle, which is controlled by the equilibrium parent temperature $T_{\text{ini}}$ from which the system is minimized. Recalling from Fig.~\ref{fig:1}(c) that the first three types of POLY systems have ductility and stability very similar to the three types of LJ systems, we notice that Fig.~\ref{fig:4}(d) and (e) are quite similar if we restrict ourselves to those data points.  This is also highlighted by the similarities in $C_{\text{min}}$ across nearly all structural indicators between ductile LJ (Fig.~\ref{fig:4}(c)) and POLY (Fig.~\ref{fig:4}(f)) systems. Taken together, these results suggest that the attractive interaction and degree of polydispersity related to the interaction potential have only a small influence on the predictive power of a given indicator. 

In contrast, the stability/ductility of the material very strongly influences the predictive power -- for almost all of the conditions studied, the predictive power increases with decreasing ductility. This can be rationalized by realizing that the system dynamics are noisy, and so when a large fraction of the system is relatively soft it is more difficult to predict which of those soft regions will fail first, consistent with Ref.~\cite{barbot2018local}.

\begin{figure}[t!]
  \includegraphics[scale=1]{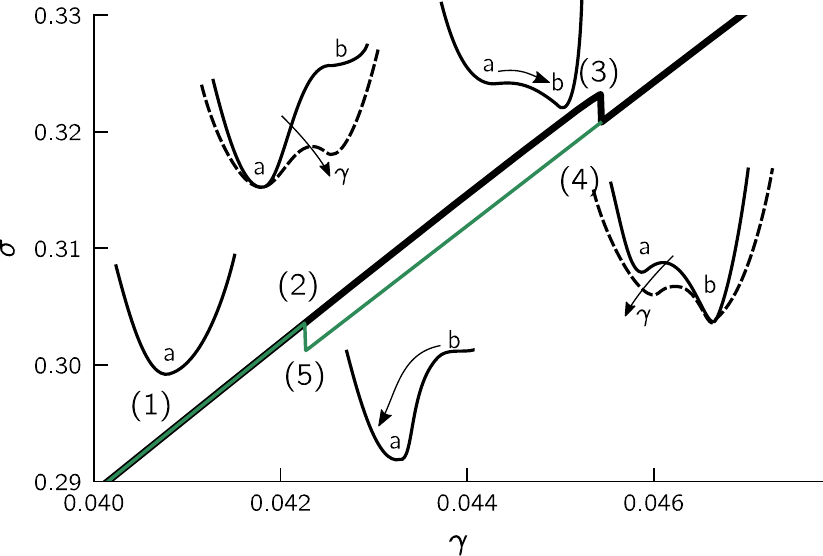}
  \caption{\textbf{Anelastic events in stable glasses.}~Example of an anelastic (reversible) plastic event in a stable glass (arrows indicate the loading direction). Rightmost sketches show the evolution of the potential energy landscape along the hysteresis. }
  \label{fig:5}
\end{figure}

On the other hand, the most brittle systems we study -- Fig.~\ref{fig:4}(g) and the right-most points in Fig.~\ref{fig:4}(e) -- buck this trend. These correspond to ultra-stable glasses that can only be formed using swap Monte Carlo. In those systems, the predictive power of vibrational modes, softness, and $\Theta$ are all lower. We also observe the same drop in Saddle Point Sampling (SPS), which highlights that the relevant saddles in the potential energy landscape (PEL) are not present at $\gamma=0$ and form during the elastic branch. One direct consequence is the reversibility of first inelastic events in very stable glasses as shown in Fig.~\ref{fig:5}  (also observed in Refs.~\cite{Jineaat6387,xu2018predicting,fsp}). During the elastic branch, the imposed loading creates a saddle that links two states (a) and (b). At the first plastic event, the system flows from (a) to (b) in the usual saddle-node bifurcation (as shown e.g. in \cite{maloney2004universal}). Shearing back the system the saddle now moves towards the state (b) as the energy barrier weakens. The recovering strain marks the location at which the original saddle was created. In contrast, the predictiveness offered by the residual plastic strength keeps increasing. This suggests that in very stable glasses, the as-cast structure is insufficient to predict the first event and that one needs access to the dynamics along the elastic branch.

\subsection{Correlation away from the as-cast glass}
One possible explanation for this observation is that (on average) the first plastic event is nearly an order of magnitude further away in strain for the most brittle materials compared to the other material preparations (Fig.~\ref{fig:1}(c)). To test this hypothesis, we study how the correlation between structure and deformation changes with the number of plastic events that have occurred since the state was prepared at $\gamma=0$. For the ductile glass, we see that the correlation function decays smoothly as the system approaches the yielding regime, identified by the gray shading (Fig.~\ref{fig:4}(h)). The correlation for some structural indicators remains above the noise floor even beyond 50 plastic events or 10\% strain, consistent with previous studies~\cite{manning2011vibrational,patinet2016connecting}. 

The fact that the correlation does not fully decay after the yielding transition suggests two possible scenarios:  either (i) soft regions that have failed during the elastic branch are still weak and fail again in the post-yielding steady state, and related (ii) some significantly harder zones in the quenched state never yielded but were simply advected with the deformation (e.g. white regions in the snapshot shown in Fig.~\ref{fig:1}(d)). These zones, rightfully measured as hard in the initial state, thus continue to participate in the correlation score.

\begin{figure}[t!]
  \includegraphics[scale=1]{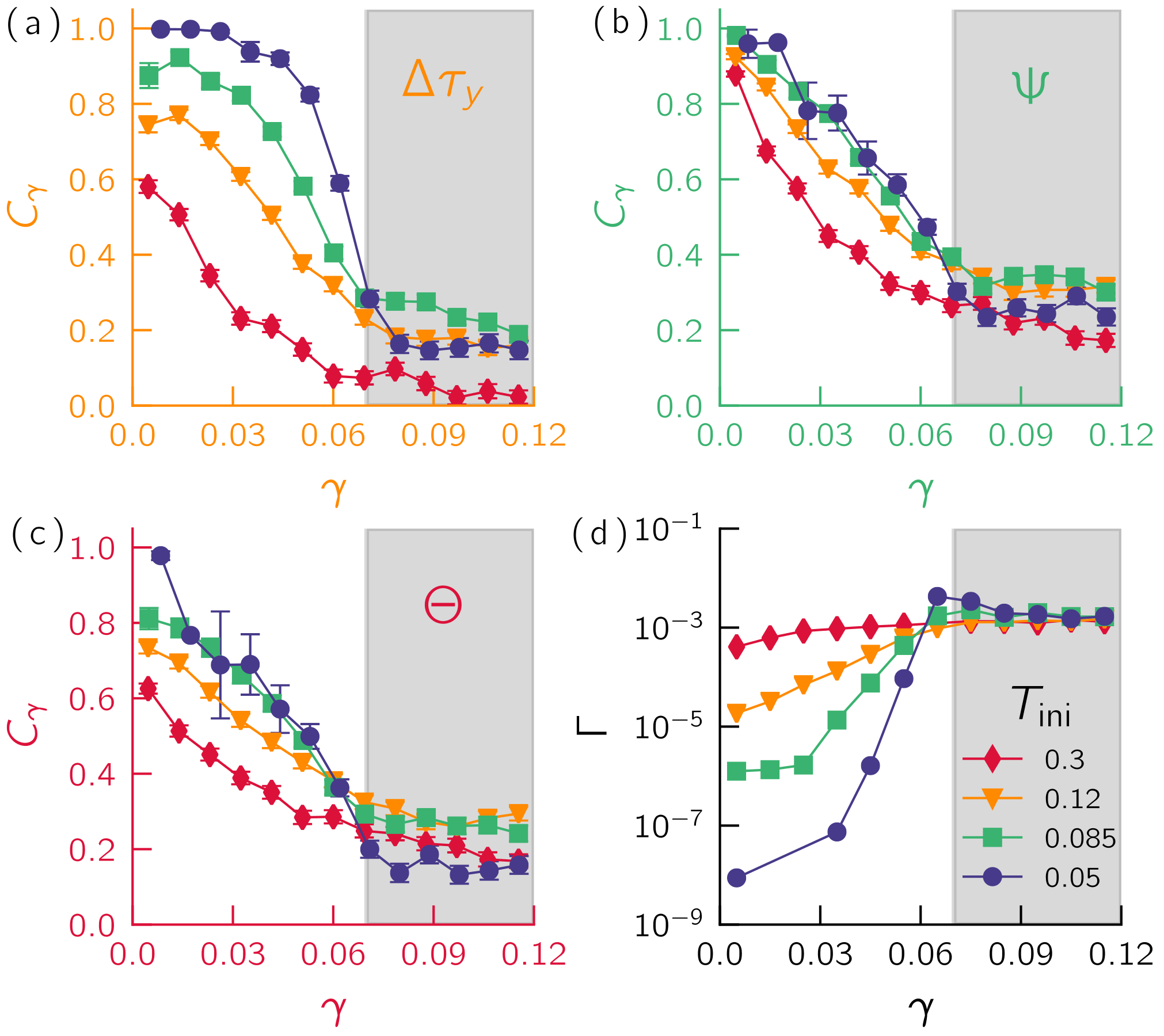}
  \caption{\textbf{Correlation decay from the as-cast glass and dissipation.}~Correlation decay $C_\gamma$ at various degrees of stability (controlled by $T_{\rm ini}$ in the POLY model) for the  residual plastic strength $\Delta\tau_y$ (a), the vibrality $\psi$ (b), and the steric bond order $\Theta$ (c). Panel (d) shows the energy dissipation density $\Gamma$ as a function of the strain $\gamma$ for the same preparation protocols.}
  \label{fig:6}
\end{figure}

In brittle glasses, with the exception of $\Delta\tau_y$, there is instead a sharp decrease in correlation after just a handful of plastic events (Fig.~\ref{fig:4}(j)). In Fig.~\ref{fig:6}, we plot the correlation between the structure at $\gamma_{\rm str}=0$ and deformation changes with the strain for various glass stabilities (POLY system). At the same time, we monitor the average energy dissipation density $\Gamma$ across the yielding transition. In the limit $\gamma\to 0$, we do observe a significant increase of the correlation as the glass becomes more stable (because of the increase of contrast due to the depletion of soft regions, see Ref.~\cite{barbot2018local}). As the plastic activity (measured by $\Gamma$) increases the correlation decays as the microscostructure is progressively reshuffled. However in stable glasses, $\Gamma$ is about four orders of magnitude lower than in poorly annealed glasses, which means that the decrease seen in $C_\gamma$ occurs in the absence of any plasticity. This suggests that significant changes to the indicators must occur along \emph{strain-reversible elastic branches}, and that the longer extent of these elastic branches is indeed the reason why, for most indicators, predictive capabilities are low in the most brittle glasses. It also explains why $\Delta\tau_y$ is far superior in this case -- because $\Delta\tau_y$ is measured only after the system yields, it has access to these transformed states far along an elastic branch. 

\begin{figure}[t!]
  \includegraphics[scale=1]{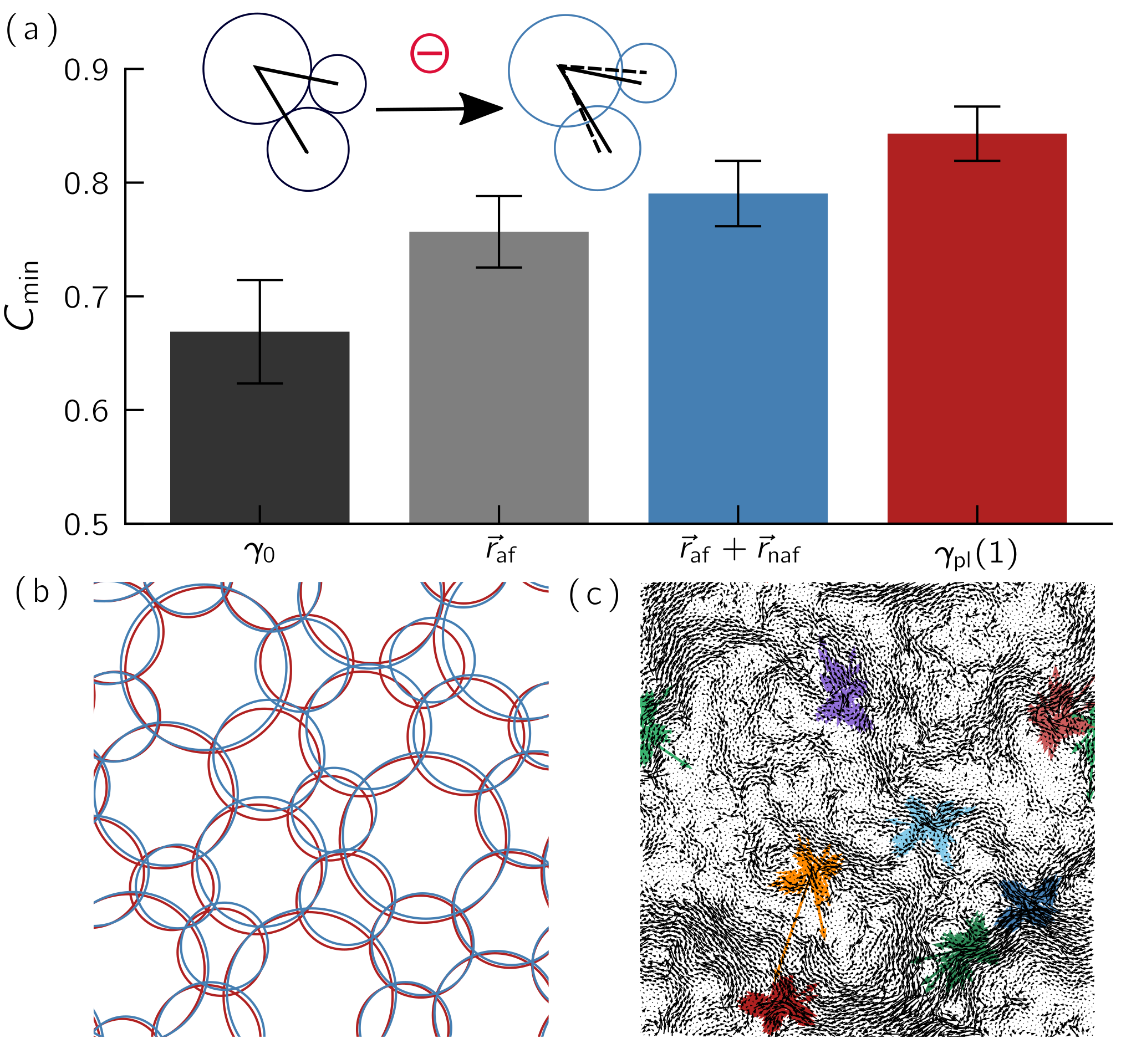}
  \caption{\textbf{Elastic branch and nonaffine motions.}~(a) Correlation $C_{\rm min}$ for the $\Theta$ indicator computed in the as-cast configuration at $\gamma=0$ (black ), affinely transformed coordinate up to the onset critical strain (gray), affinely and nonaffinely transformed coordinates (blue), and configuration at the onset strain $\gamma_{\rm pl}(1)$. The sketch illustrates the increase of frustration (internal forces) during the large elastic branch. (b) Comparison between the transformed coordinates ($\rv_{\rm af}+\rv_{\rm naf}$, blue) and positions at the onset (red). (c) Superimposition of the nonaffine velocity (black field) evaluated at $\gamma_{\rm str}=0$ and the eight softest quasilocalized excitations in the as-cast glass.}
  \label{fig:7}
\end{figure}

For a short elastic branch, one can predict at first order how the particle positions are going to evolve. In athermal dynamics, this transformation is composed of an affine $\rv_{\rm af}(\gamma)$ and a nonaffine term $\rv_{\rm naf}(\gamma)\simeq \dot{\xv}\gamma+\mathcal{O}(\gamma^2)$, where $\dot{\xv}=d\rv/d\gamma|_{\gamma_0}$ is often referred to as the nonaffine velocity, here evaluated at $\gamma=0$. To validate that it is indeed the dynamics occurring prior $\gamma_{\rm pl}(1)$ that is responsible for the drop seen in $C_{\rm min}$, we propose to recompute the $\Theta$ map in stable glasses after the affine transformation and subsequently adding the nonaffine deformation, see Fig.~\ref{fig:7}(a). We observe a clear increase of correlation. Adding both $\rv_{\rm af}$ and $\rv_{\rm naf}$ gives similar results as if one computes $\Theta$ at the onset strain $\gamma_{\rm pl}(1)$. The reason for the slightly lower correlation compared to the onset is due to higher-order non-affinities which are not encoded in $\dot{\xv}$ evaluated at $\gamma=0$, see  Fig.~\ref{fig:7}(b). Here we demonstrate that the change of geometrical frustration picked up by $\Theta$ along the elastic branch is controlled by the affine and nonaffine deformations. The latter is controlled by  soft excitations (quasilocalized modes) that couple well to the imposed nonaffine shear force. We highlight this in Fig.~\ref{fig:7}(c) by superimposing $\dot{\xv}$ computed at $\gamma=0$ with the eight lowest localized excitations extracted from our non-linear framework~\cite{gartner2016scipost}. Note that higher-order corrections in the dynamics, namely the nonaffine acceleration $\ddot{\xv}=d^2\rv/d\gamma^2$, will also be dominated by those soft excitations.

\subsection{Correlation approaching an instability}
While some of the indicators are computationally expensive and therefore can be computed only at $\gamma=0$, it is possible to compute other indicators along the strain trajectory. For these indicators, we also explore how well they capture deformation as a function of the strain until the next plastic event, $\Delta \gamma$, averaged over all plastic events. These data are shown with $C_{\Delta \gamma}=C(\gamma_{\rm pl}-\Delta\gamma,\gamma_{\rm pl})$ in Fig.~\ref{fig:4}(i) for a ductile glass, and in Fig.~\ref{fig:4}(k) for a brittle glass. In both systems, the correlation for the best indicators is nearly unity until a relative strain difference of $10^{-3}$, which is close to the average distance between plastic events for a system of this size. Beyond this strain scale, the predictive power decreases exponentially, suggesting that the system gradually loses memory of its past state over a characteristic strain scale of $10^{-1}$.

\begin{figure*}[t!]
  \includegraphics[width = \textwidth]{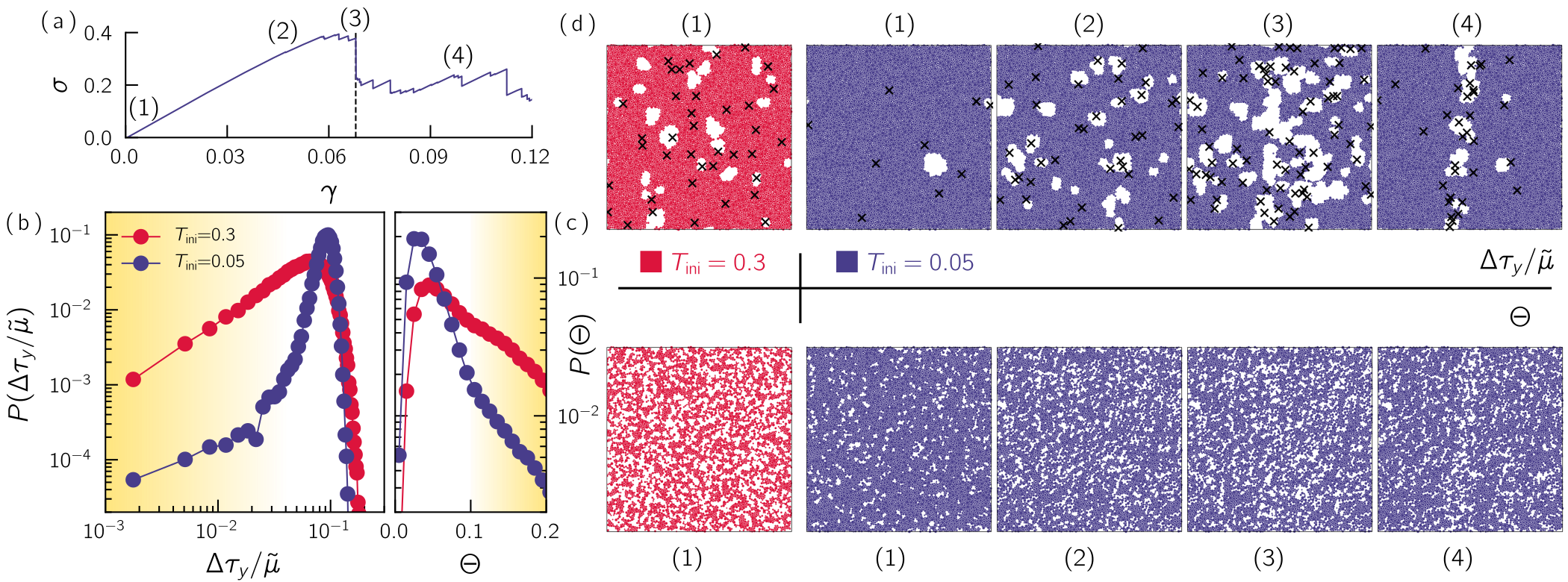} 
  \caption{\textbf{Yielding pathway.}~(a) Stress-strain curve of a brittle glass (the same sample as shown in Fig.~\ref{fig:1}(d)). (b) Probability distribution function of the (strain) distance to threshold $\Delta\tau_y/\tilde{\mu}$ at zero strain for ductile (red) and brittle (blue) glasses. (c) Probability distribution of the measure of microscopic disorder $\Theta$. (d) Snapshots highlight the spatial distribution of soft regions (white color) with particles having $\Delta\tau_y/\tilde{\mu}<3\%$ and $\Theta>0.1$, respectively. Black crosses show the location of low-energy excitations. Similar trends occur for the structural softness indicator, Fig S9.}
  \label{fig:8}
\end{figure*}

\subsection{Effects of the dimension and system size}
In this work, we focus on simple 2D glass models; revealing the degree to which our results extend to 3D and to more realistic glass models is therefore crucial. There is a sufficient body of work to suggest that lessons from analysis of 2D systems do carry over to 3D systems. Most of the indicators considered here can be computed in three dimensional systems, and some have already successfully been applied to bulk metallic glasses (BMG) such as the saddle point sampling (SPS) \cite{xu2018predicting}. We expect methods based on linear response to be highly effective in 3D as the properties of nonphononic low-frequency excitations remain unchanged with dimension \cite{kapteijns2018universal} and interaction complexity \cite{bonfanti2020universal,richard2020universality}. Supervised machine learning methods, such as the softness $S$, have been shown to be very efficient in 3D systems \cite{schoenholz2016structural,bapst2020unveiling}. On the other hand, it has been shown that e.g.~the $\Theta$ indicator is less predictive of flow in 3D, in the context of supercooled liquids' dynamics \cite{tong2018revealing}. 

Moreover, as computer glasses are inherently limited in size (the linear size of our systems is about $100$ particle diameters long), finite-size effects are expected to emerge. Although a rigorous finite-size study is beyond the scope of this paper --- due to the various preparation protocols and large number of indicators involved ---, we can still speculate on how our results would change in the thermodynamic limit based on previous studies that have focused on finite size effects. In particular, it is known that the average strain at which the first plastic event takes place scales as $N^{-1/(1+\theta)}$, with $\theta\approx2/3$ \cite{karmakar2010statistical,shang2020elastic,finite_size_intervals,episode1_geert}. As we have demonstrated that many indicators feature nearly perfect predictiveness at small strains away from instabilities (see Fig.~\ref{fig:4}), we can expect that the predictability of the first plastic event will improve with increasing system size (in particular for very stable glasses, cf.~Fig.~\ref{fig:4}(e)).

Furthermore, it has been shown~\cite{shang2020elastic} that the energy dissipation density $\Gamma$ occurring during the elastic branch remains constant with system size. Since we found that the correlation decay seen in $C_\gamma$ is mainly controlled by how much dissipation (plastic events) occurs at a given strain (see Fig.~\ref{fig:6}), we expect that our results should remain valid in the thermodynamic limit.

\section{Evolution of plastic defects across the yielding transition}
Many phenomenological models that predict plastic deformation and failure rely on largely untested assumptions about the characteristics of structural defects, such as their strain distance to threshold or density. Now we are finally in a position to begin to test some of those assumptions, by quantifying properties of our calculated structural fields and studying how different structural indicators contribute different insights into macroscopic material response. In particular, we can utilize a structural indicator to isolate regions likely to rearrange and follow their spatial evolution during deformation. Here, we propose to follow various sheared states along the stress-strain curve of a brittle glass (Fig.~\ref{fig:8}(a)).

We first focus on the initial state ($\gamma=0$). We extract an estimate for the strain distance to the next instability as $\Delta\tau_y/\tilde{\mu}$, with $\tilde{\mu}$ being the bulk shear modulus. In Fig.~\ref{fig:8}(b), we plot the distribution $P(\Delta\tau_y/\tilde{\mu})$ at zero strain, for our most ductile ($T_{\text{min}}=0.3$ (red)) and most brittle ($T_{\text{min}}=0.05$ (blue)) computationally modeled glasses. In both cases, we find a power-law tail at low $\Delta\tau_y/\tilde{\mu}$, highlighting the presence of anomalously soft regions. However, the density of regions close to a plastic rearrangement changes drastically between the ductile and brittle material (about two orders of magnitude), which is illustrated by the left-most snapshots where particles that are less than $3\%$ in strain from threshold are colored in white. We observe the same decrease in the number of low-energy excitations (extracted by localizing modes lying below the onset frequency of the power law tail in the density of states, see SM) and indicated by black crosses, consistent with other studies~\cite{lerner2017effect,wang2019low,rainone2020pinching}. We find that the purely structural indicator $\Theta$, which does not require information about the interaction between particles, is also able to resolve a decrease in the number of particles belonging to highly disordered motifs ($\Theta>0.1$), demonstrated quantitatively in the distribution $P(\Theta)$ in Fig.~\ref{fig:8}(c) and qualitatively in the left-most bottom snapshots. Since the tail of $P(\Theta)$ changes by only one order of magnitude instead of two, it incorrectly labels some regions as soft even when the distance to threshold ($\tau_y/\tilde{\mu}$) is high. Nevertheless, these data demonstrate that some purely structural methods easily accessible to experimentalists can effectively be used to sort samples with respect to their ductility. While $\Theta$ works well for these simulations with spherically symmetric interaction potentials, similar trends are seen in also in the structural softness metric $S$, as shown in Fig. S9, which can be applied to a wide range of simulations and experimental systems~\cite{cubuk2017structure}.

\begin{figure}[t!]
  \includegraphics[scale=1]{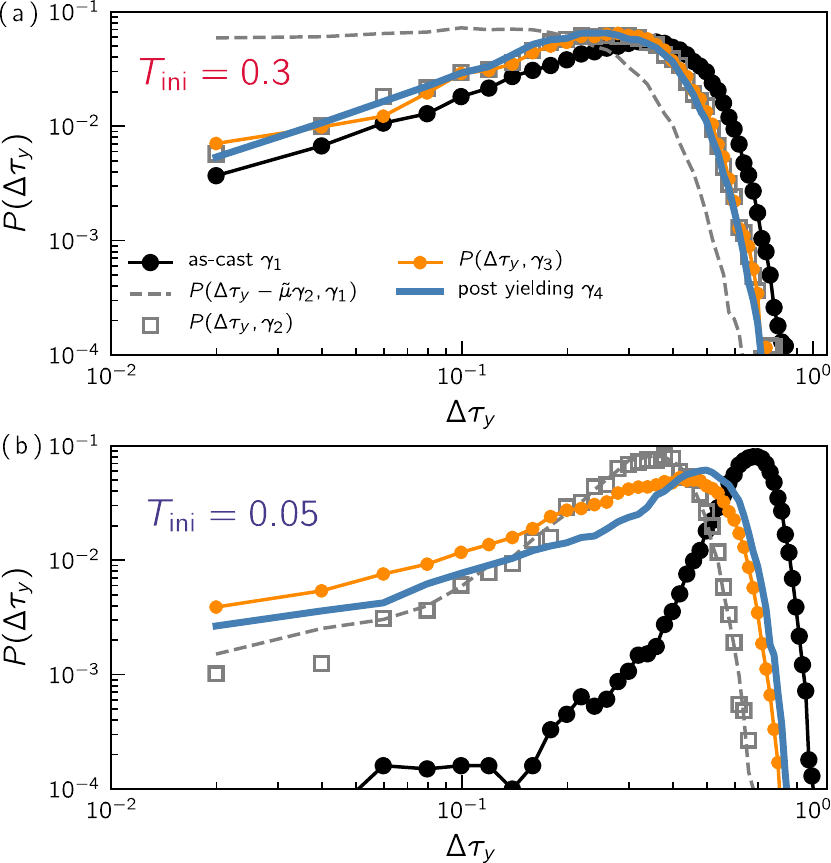} 
  \caption{\textbf{Residual plastic strength distributions as a function of strain.}~Probability distribution of the residual plastic strength $\Delta\tau_y$ at different strain corresponding to the state marked in  Fig.~\ref{fig:8}(a) for a poorly annealed (a) and very stable glasses (b). Here, $P(\Delta\tau_y,\gamma_2)$, $P(\Delta\tau_y,\gamma_3)$, and  $P(\Delta\tau_y,\gamma_4)$ is averaged over states at strain $4\%<\gamma_2<5\%$, $6\%<\gamma_3<7\%$, and $\gamma_4>10\%$, respectively.}
  \label{fig:9}
\end{figure}

We next study specific features of the residual plastic strength distributions as a function of strain, show in Fig.~\ref{fig:9}. One consequence of the fact that $P(\Delta\tau_y)$ has a significantly depleted tail for low values of $\Delta\tau_y$ in brittle glasses (shown by the black line in Fig.~\ref{fig:9}(b)) is that such glasses can be deformed up to 5\% of strain --  labeled strain point (2) in Fig.~\ref{fig:8}~and~\ref{fig:9} -- with only a minor plastic activity. At this large amount of strain, localized excitations have softened, resulting in a shift of $P(\Delta\tau_y)$ towards $\Delta\tau_y \to0$, as shown by the gray empty squares in Fig.~\ref{fig:9}(b). In other words, regions that were relatively hard at zero strain move closer to their critical threshold, in agreement with elasto-plastic models where $\Delta\tau_y$ is assumed to decrease by the elastic loading $\tilde{\mu}\gamma$.  

We test this prediction explicitly by comparing the rescaled distribution $P(\Delta\tau_y-\tilde{\mu}\gamma_2,\gamma_1)$ from the state (1) (as-cast glasses) with the distribution $P(\Delta\tau_y,\gamma_2)$ of state (2), shown by the dashed gray line and empty squares respectively in Fig.\ref{fig:9}(b). It is in good agreement with the solid gray line, suggesting that the elasto-plastic assumption works well for brittle materials. In constrast, this rescaling does not hold in ductile glasses at large strains, as already many blocks have yielded at 4-5\% of strain and thus have been redrawn (on average) at higher $\Delta\tau_y$ values. This is shown by the dashed gray lines in Fig.\ref{fig:9}(a). In addition, we note that purely structural indicators are less sensitive to this softening, as highlighted in Fig.~\ref{fig:8}(d).

\begin{figure}[t!]
  \includegraphics[scale=1]{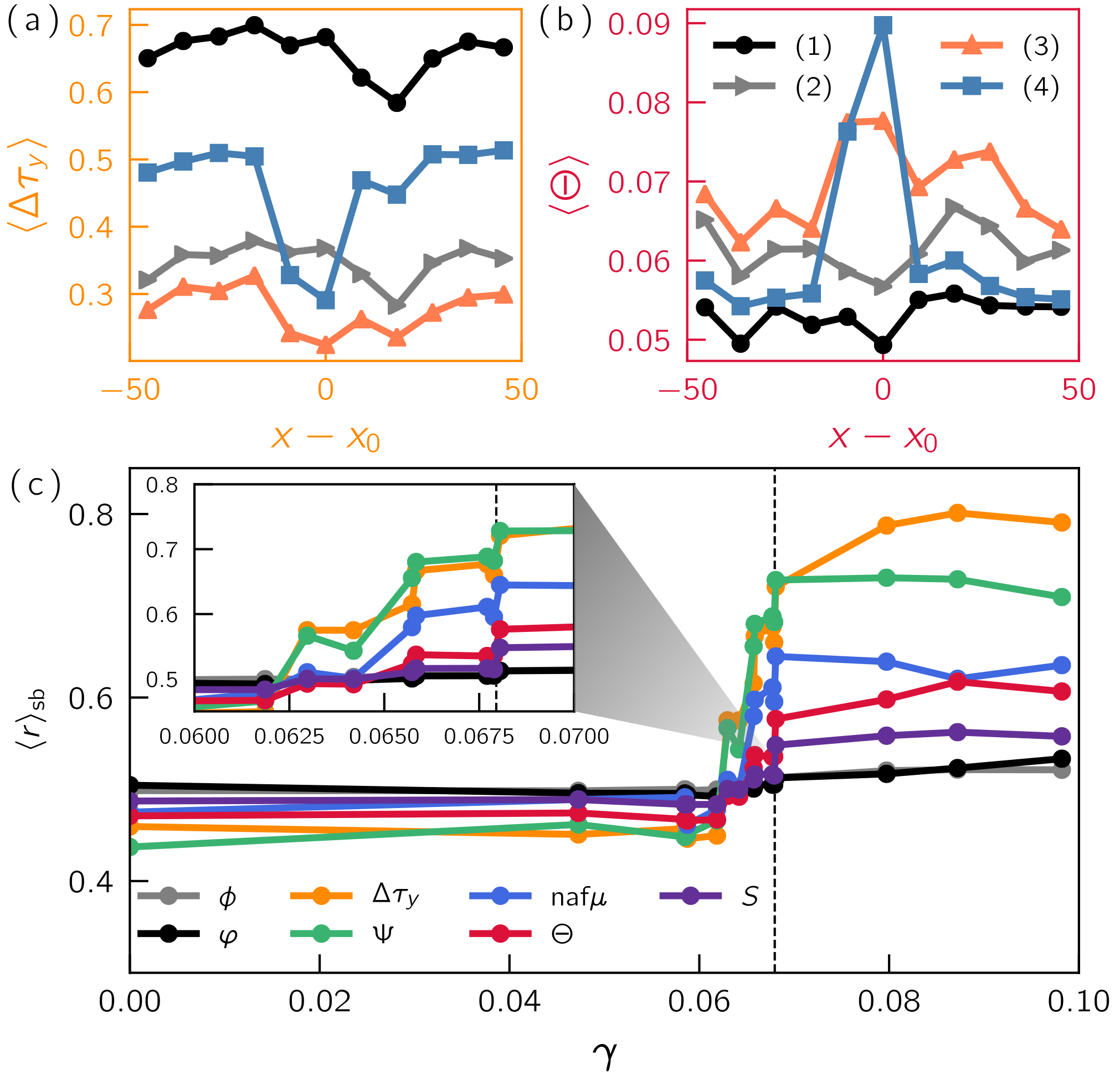} 
  \caption{\textbf{Strain softening in stable glasses.}~ Transverse average profile of the residual plastic strength $\Delta\tau_y$ (a) and steric bond order $\Theta$ (b) at the same state points shown in Fig.~\ref{fig:8}. (c) Average rank inside the shear band $\langle r\rangle_{\rm sb}$ as a function of the strain for various indicators. The vertical dashed line marks the location of the largest stress drop and the formation of a permanent shear band. Inset shows a zoom of the rank close to the yielding transition.}
  \label{fig:10}
\end{figure}

Nevertheless, as the system progresses to the yielding point -- strain state (3) -- both the ductile and brittle materials exhibit similar distributions for $P(\Delta\tau_y)$, shown by the orange lines in Fig~\ref{fig:9} (a) and (b). Both distributions exhibit a further enhancement in probability at low $\Delta \tau_y$ compared to distributions for states (1) and (2) -- a larger number of regions are closer to yielding. Of course, even though the distributions are similar, the spatial organization of soft regions could be different between the two systems, leading to their drastically different yielding behavior. To study this, we first note that in this geometry, a shear band forms oriented along the $y$ axis (state (4) in Fig.\ref{fig:8}(d)). For a given initial condition, we denote as $x_0$ the location along the $x$ axis which is the center of the shear band when it forms. We can then plot the value of various indicators, averaged over the y-direction, as a function of the distance to the center of where the shear band will form, $x-x_0$. Further, we also estimate the width of the shear band, and average the value of indicators over that width.

In Fig.~\ref{fig:10}, we plot the $y$-averaged residual strength $\langle \Delta \tau_y \rangle$ (Fig.~\ref{fig:10}(a)) and average steric order $\langle \Theta \rangle$ (Fig.~\ref{fig:10}(b)) for the same state points shown in Fig.~\ref{fig:8}(a). As discussed above, we first observe a global softening from (1) to (2) with a lower threshold $\langle \Delta \tau_y \rangle$ and higher structural disorder $\langle \Theta \rangle$. Moving from (2) to (3), we observe a clear heterogeneous and localized softening in the region where the shear band will form, in agreement with recent work \cite{barbot2020rejuvenation}. In Fig.~\ref{fig:10}(c), we provide a comparison how well different indicators capture this strain softening and thus can forecast the formation of a shear band. Here, we rank order the structural field and monitor its average inside the band $\langle r\rangle_{\rm sb}$ as a function of the strain. We find no localization (i.e. $\langle r\rangle_{\rm sb}\simeq 0.5$) up to $\gamma=6\%$ followed by a rapid increase of $\langle r\rangle_{\rm sb}$ prior to the yielding transition. Most indicators are able to capture this softening, with a better signal/noise resolution in indicators with explicit information about the force network. In contrast, we show $\langle r\rangle_{\rm sb}$ for the free volume $\phi$ and local potential energy $\varphi$ which demonstrate a very weak contrast between the inside and outside of the band at the state (3). Here, we have monitored the dynamics of a single snapshot (same one as rendered in Fig.\ref{fig:8}). Note that, averaged profiles of the local yield strength $\tau_c$ over many realizations at different strains have been performed in Ref.~\cite{barbot2020rejuvenation} and show similar results.

Finally, the system yields and reaches a transient shear-banded state (4). We observe that the majority of our structural indicators are able to locate the shear band (including $\phi$ and $\varphi$). In particular, we observe an array of low energy excitations perfectly aligned with regions close to their threshold. Moreover, $\Theta$ reveals that the microstructure inside the band is highly disordered (see Fig.~\ref{fig:8}(d) and Fig.~\ref{fig:10}(b)), reinforcing the link between local disorder, low-energy excitations, and residual plastic strength. After the large stress drop, we observe that the region outside of the band has been stabilized, so that $\langle \Delta \tau_y \rangle$ increases and $\langle \Theta \rangle$ decreases. This stabilization is due to the decrease of the applied stress after the mechanical instability and the nucleation of the shear band. This stabilization can also be seen in a shift of the median of $P(\Delta\tau_y)$ from strain points (2) to (4), plotted in Fig.~\ref{fig:9}(b). In contrast, ductile glasses show almost no variation in their distributions of residual strength and already reach the steady-state distribution at $4\%-5\%$ of strain (see Fig.~\ref{fig:9}(a)).

\section{Conclusion}
Taken together, our results demonstrate that shear-driven rearrangements in amorphous solids are deeply encoded in the structure. In ductile systems, many of the structural indicators that have been previously proposed are highly predictive of deformation at yielding and beyond. Our work indicates that two purely structural indicators (machine learning and $\Theta$), which do not require any knowledge of the interaction potential and can be immediately applied to experimental systems, perform comparably to more complicated methods in ductile solids. Another surprising observation is that the linear modes, which can be extracted from time-averaging of two-particle correlation functions in experiments~\cite{chen2010low}, outperform all other methods in ductile materials.

By analyzing ultrastable glass configurations whose first plastic instability is pushed up to 6-7\% shear strain, we show that the predictiveness of metrics based on linear response drops significantly in those systems. These results might appear counterintuitive, as the system undergoes only reversible elastic deformation. We show that no other metric --- except for the residual plastic strength metric --- can accurately predict rearrangements for such huge strain intervals. This clearly indicates that capturing quantitative information regarding the coupling strength of soft spots to the imposed loading geometry is key to high predictiveness. Future work should, therefore, focus on enhancing existing, and developing better, micromechanical-information-rich anisotropic indicators.

Using a novel nonlinear framework which allows extracting the precise location of soft quasilocalized excitations, and directly comparing them to the local yield stress map, we have firmly established that the low local yield stresses observed in some regions directly emanate from the presence of soft quasilocalized modes. Quantifying how different nearby excitations interact and self-organize will be crucial for understanding strain localization and catastrophic failure via shear banding in brittle glasses.

Finally, our work focuses on the athermal quasistatic regime, but should certainly be extended to finite temperatures and strain rates. In particular, many fundamental questions in that context have yet not been addressed, such as: (i) What is the interplay between the thermal and mechanical activation of soft spots? (ii) How do the stresses and strains generated by a single shear transformation propagate throughout the system for different imposed strain rates? Answering these questions would place us in a prime position to formulate improved theoretical frameworks and models of elastoplasticity, both on the mesoscale --- in the form of more accurate elasto-plastic lattice models --- but also on the macroscale, towards formulating observation-based constitutive relations for macroscopic elastoplasticity.


\begin{acknowledgments}
M.O thanks Hua Tong from the Shanghai Jiao Tong University for fruitful discussions and help with implementation. We are grateful for the support of the Simons Foundation for the “Cracking the Glass Problem Collaboration” Awards No. 348126 to Sid Nagel (D.R), No. 454945 (S.A.R, A. J.L), No. 454947 (P.M and M.L.M), No. 454933 (M.O and L.B). S.P acknowledges the support of French National Research Agency through the JCJC project PAMPAS under grant ANR-17-CE30-0019-01. M.L.F acknowledges support from the US National Science Foundation under Grant No. DMR-1910066/1909733. We acknowledge support from Simons Investigator Award No. 327939 (A. J.L), the University of Pennsylvania MRSEC NSF-DMR-1720530 (G.Z), and NSF-DMR-1352184 (E.S). B.X, B.S, and P.G acknowlage the funding support from  the NSF of China (Grants No.U1930402) and  the computational support from the Beijing Computational Science Research Center(CSRC). S.S acknowledges support through the J C Bose Fellowship, DST, India. E.L was supported by the Netherlands Organisation for Scientific Research (Vidi Grant 680-47-554/3259).
\end{acknowledgments}


\appendix

\section{Glass former}
\label{ap:protocol}

\subsection{System}

\subsubsection{Binary Lennard-Jones (LJ)}
The 2D binary glasses are made of $10^4$ atoms. They were obtained by quenching liquids at constant volume. The density of the system is kept constant and equals $10^4/(98.8045)^2=1.02$. We choose the composition such that the number ratio of large (L) and small (S) particles equals $N_L/N_S=(1+\sqrt{5})/4$. The two types of atoms interact via $6–12$ Lennard-Jones interatomic potentials whose parameters are: $\sigma_{SS}=2\sin(\pi/10)$, $\sigma_{LL}=2\sin(\pi/5)$, $\sigma_{SL}=1$, $\epsilon_{SS}=0.5$, $\epsilon_{LL}=0.5$, $\epsilon_{SL}=1$, $m_S=1$, $m_L=1$. The standard Lennard-Jones potentials have been slightly modified to be twice continuously differentiable functions. This is done by replacing the Lennard-Jones expression for interatomic distances greater than $R_{\text{in}}=2\sigma$ by a smooth quartic function vanishing at a cutoff distance $R_{\text{cut}}=2.5\sigma$ \cite{barbot2018local}.

\subsubsection{Polydisperse soft spheres (POLY)}
The glass-forming model consists of particles with purely repulsive soft-sphere interactions, and a continuous size polydispersity. Particle diameters, $d_i$, are randomly drawn from a distribution of the form: $f(d) = Ad^{-3}$, for $d \in [ d_{\rm min}, d_{\rm max} ]$, where $A$ is a normalization constant. The size polydispersity is quantified by $\delta=\sqrt{\langle d^2\rangle - \langle d\rangle^2}/\langle d\rangle$, where $\langle \cdots \rangle \equiv\int \mathrm{d} d f(d) (\cdots)$, and is here set to $\delta = 0.23$ by imposing $d_{\rm min} / d_{\rm max} = 0.449$. The average diameter, $\langle d\rangle$, sets the unit of length. The soft-sphere interactions are pairwise and described by an inverse power-law potential
\begin{eqnarray}
v_{ij}(r) &=& v_0 \left( \frac{d_{ij}}{r} \right)^{12} + c_0 + c_1 \left( \frac{r}{d_{ij}} \right)^2 + c_2 \left( \frac{r}{d_{ij}} \right)^4, \label{eq:soft_v} \\
d_{ij} &=& \frac{(d_i + d_j)}{2} (1-\epsilon |d_i - d_j|), \label{eq:non_additive}
\end{eqnarray}
where $v_0$ sets the unit of energy (and temperature with Boltzmann constant $k_\mathrm{B}=1$), and $\epsilon=0.2$ quantifies the degree of nonadditivity of particle diameters. We introduce $\epsilon>0$ to the model in order to suppress fractionation and thus enhance its glass-forming ability. The constants, $c_0$, $c_1$ and $c_2$, enforce a vanishing potential and the continuity of its first- and second-order derivatives of the potential at the cut-off distance
$r_{\rm cut}=1.25 d_{ij}$. We set $c_0=-1.924145348608$, $c_1=2.111062325330$, and $c_2=-0.591097451092$. We simulate a system with $N=10000$ particles within a square cell of area $V=L^2$, where $L$ is the linear box length, under periodic boundary conditions, at number density $\rho=N/V=1$. The model is the 2D version of one developed in Ref.~\cite{ninarello2017models} and subsequently studied in Ref.~\cite{ozawa2018random} for rheology.

\subsection{Glass preparation}

\subsubsection{LJ}
Three different quench protocols are considered. The first two kinds of glass are obtained after instantaneous quenches from high-temperature liquid (HTL) and equilibrated supercooled liquid (ESL) states at $T=9.62T_g^{\rm sim}$ and $T=1.13T_g^{\rm sim}$, respectively, with $T_g^{\rm sim}\simeq0.31 \epsilon_{SL}/k_B$. The last protocol consists in a gradual quench (GQ), in which temperature is continuously decreased from a liquid state, equilibrated at $1.13T_g^{\rm sim}$, to a low-temperature solid state at $0.096T_g^{\rm sim}$, over a period of $10^6 t_0$ with $t_{0}=\sigma_{SL}\sqrt{m_{S}/\epsilon_{SL}}$. All quench protocols are followed by a static relaxation via a conjugate gradient method to equilibrate the system mechanically at zero temperature. The forces on each atom are minimized up to machine precision. The same relaxation algorithm is used hereafter to study the response to mechanical loading.

\subsubsection{POLY}
Glass samples have been prepared by first equilibrating liquid configurations at a finite temperature, $T_{\rm ini}$, and then performing a rapid quench to $T=0$, the temperature at which the samples are subsequently deformed. We prepare equilibrium configurations for the polydisperse disks using swap Monte-Carlo simulations~\cite{ninarello2017models}. With probability $P_{\rm swap}=0.2$, we perform a swap move where we pick two particles at random and attempt to exchange their diameters, and with probability $1-P_{\rm swap}=0.8$, we perform conventional Monte-Carlo translational moves. To perform the quench from the obtained equilibrium configurations at $T_{\rm ini}$ down to zero temperature, we use the conjugate-gradient method given by a C++ software \cite{bochkanov2013alglib}. The preparation temperature $T_{\rm ini}$ then uniquely controls the stability of glass, and we consider a wide range of preparation temperatures, $T_{\rm ini} = 0.05-0.300$. Some representative temperatures of this model are as follows: Onset of slow dynamics, $T_{\rm onset} \approx 0.23$, mode-coupling crossover, $T_{\rm mct} \approx 0.11$, and an estimated experimental glass transition temperature, $T_g^{\rm exp} \approx 0.068$. Note that these values are slightly different from the ones presented in Ref.~\cite{berthier2019zero} due to slight difference of the number density.

\subsection{Mechanical loading}
\begin{figure}[b!]
  \includegraphics[scale=1.]{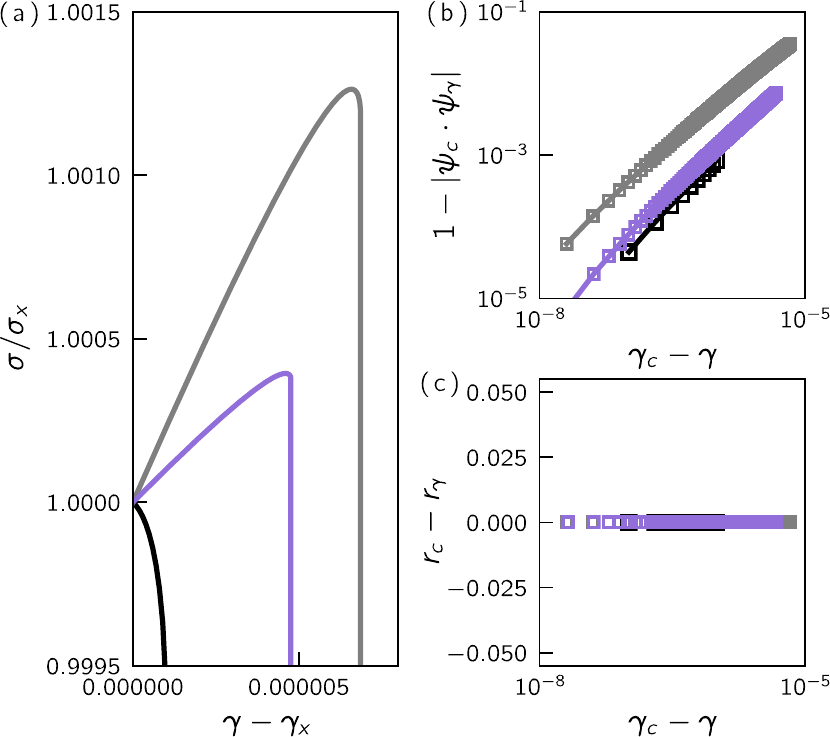}
  \caption{\textbf{Critical mode convergence.} (a) Normalized stress versus shifted strain for few saddle-node bifurcations, where $\sigma_x$ and $\gamma_x$ are the stress and strain at the onset measure with the strain step $\Delta\gamma=10^{-5}$. (b) Overlap between the critical mode $\boldsymbol{\Psi_c}$ and the mode $\boldsymbol{\Psi_\gamma}$ approaching the critical strain $\gamma_c$. (c) Rank difference between the largest component in $\boldsymbol{\Psi_c}$ and $\boldsymbol{\Psi_\gamma}$ approaching $\gamma_c$. Different colors correspond to different samples.}
  \label{fig:ap1}
\end{figure}

Beginning from a quenched unstrained configuration, the glasses are deformed in simple shear imposing Lees-Edwards boundary conditions up to $\gamma=0.12$ with an athermal quasi static method. We apply a series of deformation increments $\gamma$ to the material by moving the atom positions following an affine displacement. After each deformation increment, we relax the system to its mechanical equilibrium. In order not to miss plastic events, a sufficiently small strain increment equal to $10^{-5}$ is chosen. Plastic events are detected when the computed stress $\sigma$ decreases, a signature of mechanical instability. We have checked that the lowest mode at this onset (measured at a strain $\gamma_x$ and stress $\sigma_x$) has converged to the true critical mode $\boldsymbol{\Psi_c}$ at which the slope of the stress with respect to the strain is negative, as shown in Fig.~\ref{fig:ap1}. Fitting the critical strain $\gamma_c$ from the square root singularity of the stress (i.e. $\sigma-\sigma_c\sim\sqrt{\gamma_c-\gamma}$), we have monitored the overlap $1-|\boldsymbol{\Psi_c}\cdot\boldsymbol{\Psi_\gamma}|$ between the critical mode $\boldsymbol{\Psi_c}$ and the lowest mode mode $\boldsymbol{\Psi_\gamma}$ computed at strain $\gamma$ as a function of $\gamma_c-\gamma$, where $\gamma< \gamma_x$. In Fig.~\ref{fig:ap1}(b), we observe that the overlap at $\gamma=\gamma_x$ (the rightmost point) is already below $0.01-0.1$, meaning that the mode evaluated at $\gamma_x$ has an overlap larger than $95\%$ with the true critical mode $\boldsymbol{\Psi_c}$. More importantly for our analysis, we have extracted the rank $r_c$ of the particle having the largest $\boldsymbol{\Psi_c}$ component (core of the triggering event) and compare it to the rank $r_\gamma$ of the same particle as a function of $\gamma_c-\gamma$. We find no switch of rank from $\gamma_c$ to $\gamma_x$, see Fig.~\ref{fig:ap1}(c) where $r_c-r_\gamma=0$. We also have checked that results shown in Fig.~\ref{fig:4} remain appreciably unchanged if we simply locate loci of plasticity as the maximum of the $D^2_{\rm min}$ field over the entire avalanche. However note that in the case of large avalanches, the maximum of the $D^2_{\rm min}$ field does not necessarily correspond to the triggering event.

\section{Structural indicators}
\label{ap:indicators}

\subsection{Conventional bond orientational order $\Psi_x$}
First we consider the conventional two-dimensional bond orientational order parameters $\Psi_m^j$ for the $j$th particle, defined by
\begin{equation}
\Psi_m^j = \frac{1}{n_j} \left| \sum_{k=1}^{n_j} e^{m i \theta_{j k}}\right|,
\end{equation}
where $n_j$ is the number of nearest neighbors of the $j$th particle, and $\theta_{jk}$ is the angle between ${\bf r}_{jk}={\bf r}_k-{\bf r}_j$ and the $x$-axis.
The condition for the nearest neighbors is set to $|{\bf r}_{jk}|<x_{\rm cut} d_{jk}$, where $x_{\rm cut}$ is the first minimum of the normalized radial distribution functions. We set our cutoff after the first peak of the radial pair distribution function with $x_{\rm cut}=1.33$ and $1.5$ for the polydisperse and binary system, respectively. We use $m=2,3, \cdots, 9$, which would detect different symmetries, respectively. $m=6$ is often used to characterize hexagonal order in glassy and jamming systems~\cite{kawasaki2007correlation,schreck2011tuning}. $\Psi_6$ takes $1$ for perfectly hexagonal packings, where as $\Psi_6$ takes lower values for disordered packings.

\subsection{Generalized bond orientational order $\Theta$}
We also compute a generalized bond orientational order parameter $\Theta$ developed in Ref.~\cite{tong2018revealing}. Here we consider a central particle $j$ and its neighbor particles. The definition of the neighbors is the same as $\Psi_6$. For each pair $<kl>$ of neighbor particles next to each other, we measure the angle between ${\bf r}_{jk}$ and ${\bf r}_{jl}$, denoted as $\theta_{kl}^1$. The reference configuration with these three particles, $j$, $k$, and $l$, perfectly just in touch, with the central angle indicated as $\theta_{kl}^2$. Practically $\theta_{kl}^2$ is computed by $d_{jk}$, $d_{jl}$, and $d_{kl}$, using the cosine formula. Note that we employ nonadditive rule for $d_{jk}$ (e.g. $d_{jk}=0.5 (d_j+d_k)(1-\epsilon|d_j-d_k|)$ for the polydisperse disks), hence ``just in touch'' is achieved with respect to the nonadditive rule. Then we define the order parameter for the $j$th particle as
\begin{equation}
\Theta_j = \frac{1}{n_j} \sum_{<kl>} |\theta_{kl}^1-\theta_{kl}^2|,
\end{equation}
where $n_j$ is the number of nearest neighbors of the $j$th particle. $<kl>$ denotes the summation over all pairs of neighbors. If particles form stericallly favored, well-packed configurations (in the nonadditive sense), $\Theta$ produces smaller value, because $\theta_{kl}^1$ would be close to the reference, $\theta_{kl}^2$. Instead disordered packings generally take larger $\Theta$, since $\theta_{kl}^1$ would strongly deviate from $\theta_{kl}^2$. Thus, $\Theta$ characterizes amount of disorder, which would play the similar roles as $1-\Psi_m$, yet $\Theta$ is more sensitive order parameter for multi-components (or polydisperse) systems.

\subsection{Voronoi based metrics: $\rho$, $\phi$, $P$, and $Q$}
The local density $\rho$ and local free volume $\phi$ maps are extracted from a radical radical Voronoi tessellation by assigning to each particle a "radius" equal to its same-particle $\sigma/2$, \textit{e.g.} $\sin(\pi/10)$ for the largest particles in the LJ binary mixture. For each particle, we can define a vector $\mathBold{p}_i$ pointing from the particle center to the centroid of its Voronoi cell. For disorder packing networks, the magnitude $P(i)=|\mathBold{p}_i|$ can be anomalously large in disorder regions and thus can be used as a structural metric to quantify a local disorder. As well, one can define for each $k$ Delaunay triangle the divergence of the local interpolated field $\mathBold{p}$ as 
\begin{equation}
Q_k = (\nabla . \mathBold{p}) / (A_k/\langle A\rangle),
\end{equation}
where $A_k$ and $\langle A\rangle$ are the area of the triangle $k$ and average $\langle A_k\rangle$ over the whole packing, respectively. The local divergence $Q(i)$ is computed as the average over Delaunay triangles touching particle $i$. More details can be found in Ref.~\cite{rieser2016divergence}.

\subsection{Local excess entropy $s_2$}
To obtain an estimate of the local excess entropy $s_2$, we follow closely Ref.~\cite{piaggi2017entropy}. We first reconstruct the local radial pair distribution function $g_i(r)$ of particle $i$ as
\begin{equation}
g_i(r) = \frac{1}{A}\sum_j \frac{1}{\sqrt{2\pi\delta^2}}e^{-(r-r_{ij})^2/(2\delta^2)}.
\end{equation}
Here, the sum runs over all neighboring particle $j$ with pairwise distance $r_{ij}$, $\rho$ is the global density, and $\delta$ the standard deviation of the normalized Gaussian weight. Following previous work~\cite{williams2018experimental} for similar systems, we choose $\delta=0.12$. The particle excess entropy $s_2(i)$ follows
\begin{equation}
s_2(i) = -2\pi\rho k_B\int_0^{r_m}dr\left[g_i(r)\ln g_i(r) - g_i(r) + 1 \right],
\end{equation}
where $k_B$ is the Boltzmann constant and the cutoff $r_m$ is set to the minimum between the first and second peak of the total pair distribution function $g(r)$.

\subsection{Softness field $S$}
Following previous work \cite{schoenholz2016structural}, we use the support-vector machine (SVM) algorithm to try to fit a combination of local structural variables which best distinguishes rearranging and non-rearranging particles. Here, we aim to separate particles according to their likelihood of undergoing plastic flow. Previous works have built a training set for the SVM out of the particles which are confidently identified as rearranging (i.e. large $D_{\rm min}^2$) and nonrearranging (low $D_{\rm min}^2$ for a long time). Training only on the global maximum of the critical mode, as we test on when comparing structural indicators, would provide a very small training set, and particles near the global maximum are expected to move a large amount regardless of their structure, making them unsuitable training examples. Instead, we train our classifier on local fluctuations in the $D_{\rm min}^2$ computed on $\boldsymbol{\Psi_c}$ at each onset, which include information on both the core event and coupling with other soft regions of the system. Local extrema of $D_{\rm min}^2$ correspond to particles having a $D_{\rm min}^2$ value larger (or smaller) than all their Voronoi neighbours within a distance of two connections in the triangulation, i.e. up to the second peak of the radial distribution function. Particles corresponding to maxima and minima are labeled as $y_i=1$ (soft) and $y_i=0$ (stiff), respectively. We train an SVM as in \cite{schoenholz2016structural}, finding a linear combination of the structure functions which distinguishes local minima ($y_i=0$) from maxima ($y_i=1$). A training set of $n$ particles can be written as $\{(\Gv_i,y_i),...,(\Gv_n,y_n)\}$, where the vector $\Gv_i=(G_1,...,G_m)$ describes the local structural environment of the particle $i$ (details are provided below). Attempting to fit a deep neural network produces a higher accuracy on this training set, but is surprisingly less effective at identifying the global maximum, which reflects the fact that the training task is not exactly the same as the testing task.

The polydispersity of the POLY system necessitates unconventional choices of structural variables. In our previous work \cite{schoenholz2016structural}, most structural variables are, roughly speaking, the number of neighbor particles at a distance bin $r\sim r+dr$. Since we used a bidisperse system \cite{schoenholz2016structural}, and a small neighbor and a large neighbor can produce very different stabilizing effects to the central particle at the same $r$, we had to treat the number of small and large neighbors as two different structure functions in each bin. Unfortunately, there are infinitely many different particle sizes in the POLY system, so we cannot copy the previous approach. Our new solution is to normalize the particle distance, $r_{ij}$, by their contact distance used in the pair potential
\begin{equation}
d_{ij}=0.5\times(d_i+d_j)(1-\epsilon\times|d_i-d_j|).
\end{equation}
The number of neighbor particles with the normalized distance $r_{ij}/d_{ij}$ falling in a given bin constitutes a structure function. More precisely, we linearly spread particles into discrete bin locations, so that the structural variables are
\begin{equation}
G_m=\sum_{j} g_{m,ij}
\end{equation}
where
\begin{equation}
g_{m,ij}=
\begin{cases}
1-\frac{r_{ij}/d_{ij}-x_m}{x_{m-1}-x_m},& \text{if $x_{m-1}<r_{ij}/d_{ij}<x_{m}$,} \\
1-\frac{r_{ij}/d_{ij}-x_m}{x_{m+1}-x_m},& \text{if $x_{m}<r_{ij}/d_{ij}<x_{m+1}$,} \\
0,& \text{otherwise,}
\end{cases}
\end{equation}
and $x_m$ is the location of the $m$th radial function, given by
\begin{equation}
x_m=0.5\times1.1^{m-1}, \mbox{ } m=1,2, \cdots, 25.
\end{equation}
The linear particle spread used here is not worse than Gaussian spread used before \cite{schoenholz2016structural}, but is computationally faster. We use the diameter of the center particle, $d_i$, as one extra structure variable. We train an SVM using these 26 structure variables using regularization parameter $C=1$ and found validation accuracy $91.5\%$, $90.5\%$, $89.9\%$, and $87.5\%$; for $T_{\rm ini}=0.05$, $0.085$, $0.12$, and $0.3$, respectively. Both the training set and the validation set consists of $20000$ local-max particles and $20000$ local-min particles. Using an even larger training set significantly slows down the SVM training but produces diminishing accuracy improvements.

For the LJ system, we also normalize the separation between each pair of particles by the interaction distance $\sigma_{ij}$ between them. We then bin the normalized distances into bins of width $0.025$, between distance $0.8$ and $3$, and take the number of neighbors within each normalized distance bin as a structure function. The number of L-type neighbors and the number of S-type neighbors are treated as distinct structure functions. The diameter of the center particle is again taken as a structural variable, giving a total of $129$ structural variables. We train an SVM using these variables, again using regularization parameter $C=1$ and found validation accuracy of $93.5\%$, $92.8\%$, and $91.5\%$ for the quench protocols GQ, ESL, and HTL respectively. Both the training and validation sets consist of $60000$ local-max particle.

\subsection{Soft modes $\mathcal{M}$}
Low-frequency vibrational modes are extracted from a partial diagonialization of the harmonic dynamical matrix $\mathBold{\cal{M}}=\partial^2 U/ \partial \xv \partial \xv$. We denote the $k$th lowest eigenmode by $\mathBold{\Psi}_k$ with frequency $\omega_k=\sqrt{\kappa_k}$, where $\kappa_k$ is the corresponding eigenvalue (stiffness) of the mode. As demonstrated in many recent works~\cite{gartner2016nonlinear,zylberg2017local,schwartzman2019anisotropic}, plasticity is controlled by quasilocalized low-frequency vibrational mode. Such a mode is composed of a localized core composed of a few tens of particles and a long-ranged elastic kernel that decay as $~r^{1-d}$, with $d$ the dimension of the system. In large enough systems, these localized excitations are hybridized with plane waves (phonons). As a consequence, constructing a structural metric from stacking the norm of the $n$th lowest modes will be highly polluted by a phononic background. In Refs.~\cite{gartner2016nonlinear,zylberg2017local}, the authors have shown that one can efficiently disentangle plane waves from localized cores using the following contraction $\mathBold{\mathcal{U}^{(3)}}:\mathBold{\Psi}_k\mathBold{\Psi}_k$, where the third-order anharmonic tensor reads $\mathBold{\mathcal{U}^{(3)}}=\partial^3 U/ \partial \xv \partial \xv\partial \xv$. Each contraction of $\mathBold{\mathcal{U}^{(3)}}$ with $\mathBold{\Psi}_k$ is proportional to the mode spatial derivatives. For plane waves, such a derivative scales as the frequency $\omega$, whereas for quasilocalized excitations, it attains a characteristic value independent of frequency. As a consequence, the contribution of plane waves for low-frequency vibrational modes become negligible. In addition, the scaling of long-ranged elastic tail is suppressed and now scales as $~r^{3-3d}$ \cite{gartner2016nonlinear}. In practice, our structural indicator is computed by the weighted sum
\begin{equation}
\mathBold{\cal{M}}(i)=\sum_{k=1}^{n_k} \frac{|\mathBold{\mathcal{U}^{(3)}}:\mathBold{\Psi}_k\mathBold{\Psi}_k|_i^2}{\omega_k^2},
\end{equation}
where $|...|^2_i$ means the norm associated to the $i$th particle. We have set the number of modes $n_k$ to $512$ to optimize the prediction of plastic rearrangements.

\subsection{Local potential energy $\varphi$}
The local potential energy of the $i$th atom is computed as
\begin{equation}
\varphi(i)=\frac{1}{2}\sum_\alpha \varphi_\alpha(r_\alpha),
\end{equation}
where the sum runs over all pairs of interacting particles $\alpha=\{ij\}$ separated by a distance $r_\alpha$. The total potential energy $U$ is recovered when summing over all $\varphi(i)$.

\subsection{Local heat capacity $c_\alpha$}
The local heat capacity $c_\alpha$ associated to the interaction $\alpha$ between two particles in contact with potential energy $\varphi_\alpha$ reads
\begin{equation}
c_\alpha = \frac{\partial \varphi_\alpha}{\partial \xv \partial \xv} : {\mathBold{\cal{M}}}^{-1} - \frac{\partial \varphi_\alpha}{\partial \xv}\cdot{\mathBold{\cal{M}}}^{-1}\cdot\mathBold{\mathcal{U}^{(3)}}:{\mathBold{\cal{M}}}^{-1}.
\end{equation}
The local heat capacity $c_\alpha(i)$ of the $i$th particle is computed by summing over all interacting neighbors the absolute value $|c_\alpha|$. When performing the sum only the second (anharmonic) term $-\frac{\partial \varphi_\alpha}{\partial \xv}\cdot{\mathBold{\cal{M}}}^{-1}\cdot\mathBold{\mathcal{U}^{(3)}}:{\mathBold{\cal{M}}}^{-1}$ is kept as it is the most sensitive to low-frequency quasilocalized excitations. A more detailed description of this metric can be found in Refs.~\cite{zylberg2017local,schwartzman2019anisotropic}. The full diagonalization of the Hessian matrix $\mathBold{\cal{M}}$ is done with the Lapack library.

\subsection{Vibrality $\Psi$}
The vibrality $\Psi$ is computed following closely Ref.~\cite{tong2014order}. This indicator corresponds to the susceptibility of particle motion to infinitesimal thermal excitation in the zero temperature limit and is proportional to the well-known Debye-Waller factor. In practice, we calculate $\Psi$ as 
\begin{equation}
\Psi(i) = \sum_{k=1}^{d N-d} \frac{|\mathBold{\Psi}_k^{i}|^2}{\omega_k^2},
\end{equation}
where the sum runs over the entire set of eigenmode $\mathBold{\Psi}_k$ with frequency $\omega_k$.

\subsection{Atomic shear nonaffinity  naf$\mu$}
In athermal quasistatic deformation, the elastic constants can be derived from the second derivative of the total potential energy.  Following Maloney \textit{et al.}~\cite{maloney2004universal,barron1965second}, but in the coordinates of the normal modes, the elastic constants can be obtained as
\begin{equation}
	C_{ijkl} = \frac{1}{V}\left( \frac{\partial^2 U}{\partial \epsilon_{ij}
	\partial \epsilon_{kl} }+ \sum_m \frac{\partial^2 U}{\partial q_m \partial \epsilon_{ij}}
\cdot \frac{dq_m}{d\epsilon_{kl}}\right),
	\label{Equ:elastic_constants}
\end{equation}
where  $U$ is the potential energy, $V$ is the volume, and $q_m$ is the $m^{th}$ coordinate of the eigenbasis corresponding to the Hessian matrix ($\frac{\partial^2 U}{\partial r_{0i}\partial r_{0j}}$).  The first term of Eq.~\ref{Equ:elastic_constants}, often called Born term, is the contribution due to affine displacement, while the second term represents the contribution from nonaffine relaxation in each normal mode.

Cheng \textit{et al.}~\cite{cheng2009configurational} once observed that the nonaffine modulus of a system is much more sensitive to structural stability than the affine modulus (Born term). Considering that the stress of the system can be expressed as $\sigma_{ij}=\frac{1}{V}\frac{\partial U}{\partial\epsilon_{ij}}$ and  $\frac{dq_m}{d\epsilon_{kl}}=-\frac{1}{\lambda_m}\frac{\partial\sigma_{kl}}{\partial q_m}$~\cite{maloney2004universal}, where $\lambda_m$ is the eigenvalue of $m^{th}$ normal mode, the nonaffine contribution to the modulus from the $m^{th}$ normal mode can be rewritten as 
\begin{equation}
\tilde{C}_{ijkl,m} = 
  -\frac{V}{\lambda_m}\frac{\partial\sigma_{ij}}{\partial q_m}
  \frac{\partial\sigma_{kl}}{\partial q_m}.
	\label{Equ:nonaffine_modulus}
\end{equation}
We note that the nonaffine modulus contribution $\tilde{C}_{ijkl,m}$ is always negative and the nonaffine modulus of system can be written as $\tilde{C}_{ijkl}=\sum_m \tilde{C}_{ijkl,m}$. 

For one mode, different atoms often contribute differently. We may express  the normalized eigenvector as $\bm{\Psi}_m = \sum_{n,\alpha} c_{mn\alpha}\bm{e}_{n\alpha}$, where $\bm{e}_{n\alpha}$ is a unit vector corresponding to the displacement of $n^{th}$ atom in the $\alpha (=x, y, \mathrm{or}\, z)$ direction, and $c_{mn\alpha}$ is the projection of the $m^{th}$ eigenvector on the basis of Cartesian coordinates $\bm{e}_{n\alpha}$. Summing the  contributions from different modes, the atomic nonaffinity that quantifies the atomic nonaffine modulus contribution of each individual atom can be obtained as
\begin{equation}
\hat{C}_{ijkl,n} = \sum_{m,\alpha}-\frac{V}{\lambda_m}\frac{\partial \sigma_{ij}}{\partial q_m}\frac{\partial \sigma_{kl}}{\partial q_m}c_{mn\alpha}^2.
	\label{Equ:atomic_nonaffine_modulus}
\end{equation}
In this way, the nonaffine modulus of a system can also be written as a sum of the atomic nonaffinity of each atom as $\tilde{C}_{ijkl}=\sum_n\hat{C}_{ijkl,n}$. Since the nonaffine contribution to the modulus must converge to a finite value in the thermodynamic limit, the atomic nonaffinity must scale like $\frac{1}{N}$ ($N$ is number of atoms in the system), generally being smaller for atoms in large system.

To understand the atomic nonaffinity, we will simplify the above tensor expression to the case specific to the shear protocol, which is the most common deformation protocol of interest since local plastic  rearrangements are typically shearlike~\cite{argon1979plastic,falk1998dynamics}. We focus on the atomic shear nonaffinity, which is the shear component of the atomic nonaffinity and depends on the specified shear direction. Based on Eq.~\ref{Equ:atomic_nonaffine_modulus}, the atomic shear nonaffinity can be obtained as 
\begin{equation}
	\hat{G}_{n} = \sum_{m,\alpha}-\frac{V}{\lambda_m}
	\left( \frac{\partial \tau}{\partial q_m} \right)^2 c_{mn\alpha}^2,
	\label{Equ:ANSMC}
\end{equation}
where $\partial\tau/\partial q_m$ is the derivative of shear stress with respect to coordinate $q_m$ along the $m^{th}$ mode.

\subsection{Local shear modulus $\mu$ and local thermal expansion $\alpha$}
We calculated the local shear modulus ($\mu$)\cite{tsamados2009local} and local thermal expansion ($\alpha$) \cite{shang2018role} using a coarse-grained method originally proposed by  Goldhirsch and Goldenberg\cite{goldhirsch2002microscopic}, which connected discrete atomic position with continuum fields.

First, we defined the local coarse-grained displacement fielding $\mathbf{u}(\mathbf{r},t)$ from atomic displacement as
 \begin{equation}
   \mathbf{u}(\mathbf{r},t) \equiv \frac{\sum_{i}m_{i}\mathbf{u}_{i}(t)\phi[|\mathbf{r}-\mathbf{r}_{i}(t)|]}
  {\sum_{j}m_{j}\phi[|\mathbf{r}-\mathbf{r}_{j}(t)|]}
  \label{eqn:1}
\end{equation}
where $\mathbf{u}_{i}(t)$ is the displacement of atom $i$ at time $t$, starting from a reference position, and $\phi(x)$ is the coarse-grained function, here we choose a coarse-grained function  $\phi(r) = \frac{1}{A}[1-2(r/r_c)^4+(r/r_c)^8]$ for $r<r_c$ and 0 for otherwise\cite{barbot2018local}, with $r_c$ the coarse-grained scale and $A=8/15\pi r_c^2$. As prescribed by Tsamados \textit{et al.}\cite{tsamados2009local}, we choose $r_c=5 \sigma$ to maintain the validity of linear elasticity and heterogeneity in the mesoscale, where $\sigma$ is the atomic diameter, note that the coarse-grained size within a certain range would not change the qualitative conclusion as proved in \cite{shang2018role}.

Then based on Eq.~\ref{eqn:1}, one can obtain the local strain field $\epsilon_{ij}$ under the linear elastic assumption:
\begin{equation}
\epsilon_{\alpha \beta}(\mathbf{r})=\frac{1}{2}\left(\frac{\partial{u_\alpha(\mathbf{r})}}{\partial{x_\beta}}+\frac{\partial{u_\beta(\mathbf{r})}}{\partial{x_\alpha}}\right) 
\end{equation}

And the local stress field $\sigma_{\alpha \beta}(\mathbf{r})$ can be obtained by atomic interaction as \cite{tsamados2009local,goldhirsch2002microscopic}.
\begin{equation}
\sigma_{\alpha \beta}(\mathbf{r},t) =-\frac{1}{2} \sum_{i} \sum_{j\neq i} \frac{\partial \Psi}{\partial r_{ij}^{\alpha}} r_{ij}^{\beta} \int_{0}^{1}ds \phi(\mathbf{r}-\mathbf{r}_{i}+s\mathbf{r}_{ij})
\end{equation}
where $\Psi$ is atomic total energy. And for the small deformation, the local shear modulus $\mu$ is the response of local shear stress by the change of local shear strain using athermal quasistatic simple shear:
\begin{equation}
\mu(\mathbf{r})=\frac{\partial \sigma_{xy}(\mathbf{r})}{\partial \epsilon_{xy}(\mathbf{r})}
\label{eqn:2}
\end{equation}

And we can define the local thermal expansion $\Gamma(\mathbf{r})$ from the coarse-grained volumetric strain $\epsilon_{v}(\mathbf{r})$ caused by temperature. 
\begin{equation}
\alpha(\mathbf{r}) = \frac{\epsilon_v(\mathbf{r})}{\Delta T} \big |_{\rho}
\label{eqn:3}
\end{equation}
We calculated the local thermal expansion at constant number density in the 2D polydisperse model, we reheat the sample from inherent structure to 0.01 $T_\text{MCT}$ ($\Delta T=0.01 T_
\text{MCT}$) with the NVT ensemble to calculate local thermal expansion.

\subsection{Non-affine velocity $\dot{x}$}
The nonaffine velocity field $\dot{\xv}$ is computed from solving
\begin{equation}
\mathBold{\cal{M}}\cdot\dot{\xv} = -\frac{\partial^2 U}{\partial\xv\partial\gamma}.
\end{equation}
This displacement field is nothing than the linear response of the system to the shear force $-\frac{\partial^2 U}{\partial\xv\partial\gamma}$. This indicator is dominated by the lowest harmonic eigenmodes, which include both a phononic background and quasilocalized excitations. As done in the soft modes indicator $\cal M$, we disentangle plane waves and elastic kernels from localized cores by computing the contraction $|\mathBold{\mathcal{U}^{(3)}}:\dot{\xv}\dot{\xv}|$, giving us a unique scalar value for each particle.

\subsection{Non-linear modes $\pi$}
Following Refs.~\cite{gartner2016nonlinear,gartner2016scipost,kapteijns2019nonlinear}, we extract nonlinear modes by finding iteratively solution $\bm{\hat \pi}$ of the nonlinear equation
\begin{equation}
\mathBold{\cal{M}}\cdot \bm{\hat \pi}= \frac{ \mathBold{\cal{M}}:\bm{\hat \pi}\bm{\hat \pi} }{ \mathBold{\mathcal{U}^{(n)}} \sbullet \bm{\hat \pi^{(n)}} } \mathBold{\mathcal{U}^{(n)}}  \sbullet \bm{\hat \pi^{(n-1)}},
\end{equation}
where $\mathBold{\mathcal{U}^{(n)}}$ is the rank-$n$ tensor of derivatives of the potential energy and $ \sbullet \bm{\hat \pi^{(n)}}$ denotes a contraction over $n$ instances of the mode $\bm{\hat \pi}$. In practice, we have used $n=4$ and $n=3$ for the POLY and LJ system, respectively. The latter system having a potential not as smooth as the polydisperse model and for which $\mathBold{\mathcal{U}^{(4)}}$ is ill-defined.

In order to probe homogeneously in the system modes with both a low stiffness and a high coupling with the imposed deformation, we decompose our system into a cubic grid with cell size $l=4\sigma$. In each cell, we pick the pair $\alpha$ that has the largest $|\frac{\partial\varphi_\alpha}{\partial\xv_\alpha}\cdot \dot{\xv}_\alpha|$ value (where $\varphi_\alpha$ is the potential energy of the contact $\alpha=\{ij\}$, with $\xv_\alpha=\xv_j-\xv_i$ and  $\dot{\xv}_\alpha=\dot{\xv}_j-\dot{\xv}_i$) and compute the associated dipole response $\vec{d}_\alpha$ from solving
\begin{equation}
\mathBold{\cal{M}}\cdot\bm{d_\alpha} = \bm{f_\alpha},
\end{equation}
with $\bm{f_\alpha}=\partial \varphi_\alpha /\partial \xv$; see, for example, Ref.~\cite{lerner2018characteristic}. In an upcoming paper series~\cite{kapteijns2019nonlinear}, we demonstrate that the dipole response is an excellent starting guess to efficiently find the solution $\hat \pi$.

Glassy nonlinear modes strongly overlap in space with quasilocalized modes present in the harmonic approximation but without the phononic background~\cite{lerner2018characteristic}. Cores are still decorated with a quadrupole like elastic kernel. This long-ranged decay can be suppressed by computing the contraction $\mathBold{\mathcal{U}^{(3)}}:\bm{\hat \pi}\bm{\hat \pi}$. Finally, the structural metric $\pi$ is extracted  by summing over the $n_k$ different modes found during the mapping procedure,
\begin{equation}
\pi(i)=\sum_{k=1}^{n_k} \frac{|\mathBold{\mathcal{U}^{(3)}}:\bm{\hat \pi}_k\bm{\hat \pi}_k|_i^2}{\kappa_k},
\end{equation}
where $\kappa_k$ is the stiffness of the $k$th mode.

\subsection{Saddle Point Sampling}
A complete description of the Saddle Point Sampling can be found in Refs.~\cite{xu2017strain,xu2018predicting}. For one system with volume $V$ at initial shear strain $\gamma_0$, knowing the activation energy of a plastic event $Q_0$ and the shear stress difference between the system at a saddle point and at its initial state $\Delta \tau=\tau(x_s,\gamma_0)-\tau(0,\gamma_0)$, one can predict the triggering strain of this plastic event as
\begin{equation}
\Delta \gamma_c = \gamma_c - \gamma_0 = -\frac{3Q_0}{2V\Delta\tau_0}
\end{equation}

For one local region, if the triggering strain of all possible events are obtained, then we define the local yield strain to be the lowest one of those possible events,
\begin{equation}
\Delta \gamma_c = \min(\Delta\gamma^i),
\end{equation}
where $i$ loops over all the possible local events.

We use activation relaxation technique nouveau (ARTn) \cite{mousseau2012activation} to harvest the plastic events for each local region, with a push back strategy (mentioned in Ref.~\cite{rodney2009distribution}) to confirm each saddle point is connected to initial minimum. The activation of ARTn was initiated by imposing a random displacement to the local cluster centered on a chosen atom with a radius of $2\sigma$. A force tolerance of $5\times10^{-3}\epsilon/\sigma$ is used for converging to the saddle points. Multiple activations were attempted until five events were found for each local cluster.

\subsection{Residual plastic strength $\Delta\tau_y$}
We compute the local yield stress fields whose method is presented extensively in Refs.~\cite{patinet2016connecting,barbot2018local}. It gives access to a relevant mechanical quantity, i.e., a local slip threshold, in a nonperturbative way, over a well-defined length scale and for arbitrary loading directions. Local stress thresholds appear to be a very sensitive probe of the preparation of the glass, the anisotropy induced by plastic deformation, and the rejuvenation process~\cite{patinet2016connecting,barbot2020rejuvenation}.

This method consists in locally shearing a circular region of radius $R_{free}=5$ using the Athermal Quasi-Static method which are embedded in a shell where atoms are constrained to affine strain. It thus forces plastic rearrangements to take place in the central relaxed zone. The local yield stress $\tau_{c}$, and the associated critical deformation $\epsilon_{c}$, are computed at the critical state before the first shear stress drop in the loading direction $\alpha$. Strictly speaking, rather than calculate a threshold, we are interested in a more effective quantity to correlate structure and plastic activity which is the residual plastic strength~\cite{lemaitre2006dynamical,lin2014density} $\Delta\tau_{c}(\alpha)=\tau_{c}(\alpha)-\tau_{0}(\alpha)$. It corresponds to the amount of stress necessary to trigger an instability, where $\tau_{0}(\alpha)$ is the prestress in the probed area. On this scale, glasses are heterogeneous and anisotropic. For an external load in the $\alpha_{l}$ direction, the effective threshold thus corresponds to the smallest $\Delta\tau_{c}(\alpha)$ projected in this direction, which writes $\Delta\tau_{y}(\alpha_{l})=\min_{\alpha}\Delta\tau_{c}(\alpha)/\cos(2[\alpha-\alpha_{l}])$ with $|\alpha-\alpha_{l}|<45^{\circ}$. We compute the local yield stresses on a regular square grid of lattice parameter $R_{sampling}\approx 2.5\sigma$ with $R_{free}=5\sigma$ every $\Delta\alpha=10^{\circ}$, i.e. in 18 different directions $\alpha$. These parameters optimize the correlation between $\Delta\tau_{y}(\alpha_{l})$ and plastic activity~\cite{patinet2016connecting,barbot2018local}. In order to define a field of residual plastic strength per atom, $\Delta\tau_{y}$ is then evaluated by assigning to each atom the smallest value of the thresholds calculated at the grid points located at distances less than $R_{free}$.

%

\end{document}